\documentclass[a4paper,11pt]{article}
\usepackage{jcappub}
\usepackage[T1]{fontenc}
\usepackage[caption=false]{subfig}
\usepackage{hyperref}
\usepackage{amsmath}
\usepackage{mathtools}
\usepackage{tablefootnote}
\usepackage{soul}
\usepackage{ulem}
\usepackage{color}
\usepackage{float}
\usepackage{multirow}
\usepackage{comment}
\usepackage[usenames,dvipsnames,svgnames,table]{xcolor}

\title{Exploring galaxies-gravitational waves cross-correlations as an astrophysical probe}

\author[a,b,c]{Giulio Scelfo,}
\emailAdd{giulio.scelfo@sissa.it}

\author[a,b,c]{Lumen Boco,}
\emailAdd{lboco@sissa.it}

\author[a,b,c,d]{Andrea Lapi,}
\emailAdd{lapi@sissa.it}

\author[a,b,c,d]{Matteo Viel}
\emailAdd{viel@sissa.it}

\affiliation[a]{SISSA, Via Bonomea 265, 34136 Trieste, Italy}
\affiliation[b]{INFN, Sezione di Trieste, Via Bonomea 265, 34136 Trieste, Italy}
\affiliation[c]{IFPU, Institute for Fundamental Physics of the Universe, via Beirut 2, 34151, Trieste, Italy}
\affiliation[d]{INAF/OATS, Osservatorio Astronomico di Trieste, via Tiepolo 11, I-34143 Trieste, Italy}

\abstract{Gravitational waves astronomy has opened a new opportunity to study the Universe. Full exploitation of this window can especially be provided by combining data coming from gravitational waves experiments with luminous tracers of the Large Scale Structure, like galaxies. In this work we investigate the cross-correlation signal between gravitational waves resolved events, as detected by the Einstein Telescope, and actively star-forming galaxies. The galaxies distribution is computed through their UV and IR luminosity functions and the gravitational waves events, assumed to be of stellar origin, are self-consistently computed from the aforementioned galaxies distribution. We provide a state-of-the-art treatment both on the astrophysical side, taking into account the impact of the star formation and chemical evolution histories of galaxies, and in computing the cross-correlation signal, for which we include lensing and relativistic effects. We find that the measured cross-correlation signal can be sufficiently strong to overcome the noise and provide a clear signal. As a possible application of this methodology, we consider a proof-of-concept case in which we aim at discriminating a metallicity dependence on the compact objects merger efficiency against a reference case with no metallicity dependence. When considering galaxies with a Star Formation Rate $\psi > 10 \: M_{\odot} /\rm{yr}$, a Signal-to-Noise ratio around a value of 2-4 is gained after a decade of observation time, depending on the observed fraction of the sky. This formalism can be exploited as an astrophysical probe and could potentially allow to test and compare different astrophysical scenarios.}

\begin{document}
\maketitle

\section{Introduction}\label{sec:intro}

After the first detection of a Gravitational Wave (GW) signal was announced, originating from the merger of Binary Black Holes (BBH) of a total mass $M_{\mathrm{tot}} \sim 60 M_\odot$~\cite{abbott:firstligodetection, abbott:firstligodetectionproperties}, the era of GW astronomy began. Its groundbreaking importance comes from the fact that it opened a completely new way to observe the cosmos, using an observable that could not be exploited before. Among the newly opened directions, the birth of GW astronomy also led to new possibilities in the multimessenger field of tracers cross-correlations.

The study of cross-correlations between distinct tracers is not new. Indeed, several studies regarding e.g., correlations between the Large Scale Structure (LSS) and the Cosmic Microwave Background (see e.g., refs.~\cite{ho:correlation, hirata:correlation, Bianchini:2014dla, Bianchini:2015fiw, Bianchini:2015yly, Mukherjee:gwxcmb}), neutrinos (see e.g., ref.~\cite{fang:cross}) or among different LSS tracers (see e.g., refs.~\cite{Martinez:cross,Jain:cross,Yang:cross,Paech:cross}) have been performed. For what concerns the specific possibility to cross-correlate GW signals with LSS tracers, various works on different applications have been made, such as the investigation of the origin of merging BBHs~\cite{raccanelli:pbhprogenitors, Scelfo_2018}, the study of anisotropies of the number density and luminosity distances of compact binaries~\cite{namikawa:cross_ng} and of the stochastic GW background~\cite{alonso:cross, bertacca:sgwb}, the investigation of the GW bias~\cite{Calore:crosscorrelating}, the possibility of alternatives to General Relativity~\cite{camera:gwlensing} and several others (see e.g., refs.~\cite{Mukherjee:gwxlss1, Mukherjee:gwxlss2}).

In this paper we extend the work addressed by the community in the $\mathrm{GW} \times \mathrm{LSS}$ area studying the measurable cross-correlation signal with a refined description of both the LSS and GW tracers. Regarding the characterization of the LSS tracers we make use of actively star-forming galaxies. Observations of the last decade, with the advent of high redshift far-IR/sub-mm surveys, have helped in robustly characterizing the galaxies luminosity functions, from which the Star Formation Rate (SFR) $\psi$ can be derived, allowing us to have a rather solid statistics of actively star-forming galaxies (for a more detailed discussion see section \ref{sec:SFR} and references therein). Therefore, we exploit the galaxy SFR to organize different galaxy types in different SFR bins. In this way we can look at the contribution to the cross-correlation signal coming from galaxies with different star formation activities. 

On the other hand, we model the redshift distributions of the detected GW signals, coming from all types of merging compact objects COs (BH-BH, BH-NS, NS-NS, where NS stands for Neutron Stars), by convolving a detector sensitivity curve with the intrinsic merging rates. We self-consistently derive the latters from the aforementioned galaxies distribution. Thus, the two tracers considered here are not coming from different and independent sources: we are looking at the same objects (galaxies) but through different messengers (light and GWs). However, we need to take into account that the GWs distribution does not only depend on the SFR of the galaxies. Both the number of merging compact objects and their chirp mass (affecting the GW signal detectability) can strongly depend also on the environmental conditions in which the binary forms and evolves. For these reasons, in this work we use a refined determination of the COs merging rates, following reference~\cite{boco+19}, where a metallicity distribution is associated to each single galaxy through a chemical evolution model. Once the intrinsic merging CO distribution is computed, we convolve it with the sensitivity curve of the future third generation GW observatory Einstein Telescope (ET)~\cite{Sathyaprakash:ET} to get the detected GW events redshift distribution. 

These characterizations altogether lead to our forecast on the cross-correlation signal that can be obtained by realistically modeling these two types of tracers, especially awaiting the soon-to-come GWs detections from third generation observatories. We perform here both a tomographic and a non tomographic approach.

As a possible application of this cross-correlation formalism, we explore the idea of testing different astrophysical scenarios, which predict different GWs distributions, clustering and other specifics. We consider an exemplificative proof-of-concept case, in which we test whether metallicity dependencies on the COs merging efficiency can be detected and distinguished with respect to a benchmark case with no metallicty dependence. Looking at the distribution and clustering of these two tracers can be a promising tool to discriminate between different features imprinted by several astrophysical mechanisms.

This work is structured as follows: in section \ref{sec:methods} we provide the mathematical background to describe the cross-correlation signal, given by the number counts angular power spectra $C_\ell$'s; in section \ref{section:tracers} we describe the tracers considered and the theoretical background behind the physical quantities characterizing them; in section \ref{sec:Cl_SN} we present estimates on the cross-correlation signals through Signal-to-Noise computations; in section \ref{sec:discerning_scenarios} we describe how to potentially distinguish astrophysical features from different models and predict its viability using a test scenario; in section \ref{sec:conclusions} we draw our conclusions.

\section{Cross-correlation formalism}
\label{sec:methods}

We describe the cross-correlation between two tracers by working in the harmonic space, considering as observable the number counts angular power spectrum $C_\ell$. The multipole number $\ell$ relates to the angular resolution $\theta$ as $\ell\sim 180^o/\theta$ (see e.g., refs.~\cite{heavens:rsdspherical, szalay:wideangle, matsubara:wideangle, papai:wideangle, raccanelli:wideangle, bertacca:wideangle, raccanelli:wideangleII, dai:tamwaves, yoo:wideangle, blake:wideangle, taruya:wideangle} for works about giving up the flat sky approximation and the advantages of working in the harmonic space). What follows is valid both in the case when tomographic maps of the two tracers are available and in the case in which tomography is not performed, whereas the latter case can simply be seen as the first one reduced to a single redshift bin.

Being $X$ and $Y$ the two tracers, we can write the relation between the \textit{observed} angular power spectrum $\tilde{C}^{XY}_\ell(z_i,z_j)$ (obtained cross-correlating tracer $X$ in redshift bin $z_i$ with tracer $Y$ in bin $z_j$) and the harmonic coefficients $a_{\ell m}$ as:
\begin{equation}
\langle a^X_{\ell m}(z_i) a^{Y^*}_{\ell'm'}(z_j)\rangle = \delta_{\ell \ell'} \delta_{mm'}\tilde{C}^{XY}_\ell(z_i,z_j), 
\end{equation}
where $\delta$ is the Kronecker delta. The observed harmonic coefficients $a_{\ell m}^X(z_i)$ are given by the sum of the partial wave coefficients of the signal and of the noise:
\begin{equation}
a_{\ell m}^X(z_i) = s_{\ell m}^X(z_i) + n^X_{\ell m}(z_i).
\end{equation}
The observed angular power spectra read as
\begin{equation}
 \tilde{C}^{XY}_\ell(z_i,z_j) = C^{XY}_\ell(z_i,z_j) + \delta_{XY} \delta_{ij} \mathcal{N}^X_\ell(z_i).
\label{eq:tilde_Cls}
\end{equation}
The angular power spectrum is directly obtained from the signal wave coefficients as~\cite{raccanelli:crosscorrelation, pullen:crosscorrelation}
\begin{equation}
\langle s^X_{\ell m}(z_i) s^{Y^*}_{\ell' m'}(z_j) \rangle = \delta_{\ell\ell'} \delta_{mm'} C^{XY}_\ell(z_i,z_j).
\end{equation} 
We construct the noise angular power spectrum from the shot noise $\mathcal{N}^X_\ell(z_i)$ (inversely proportional to the number of sources per steradian), assuming no other sources of error and no correlation between noise terms of different experiments and $z$ bins. 
The expectation value of the noise can then be written as
\begin{equation}
\langle n^X_{\ell m}(z_i) n^{Y^*}_{\ell' m'}(z_j) \rangle = \delta_{\ell\ell'} \delta_{mm'} \delta_{XY} \delta_{ij} \mathcal{N}^X_\ell(z_i).
\label{noise}
\end{equation}
Assuming signal and noise as statistically independent, one can write
\begin{equation}
\langle s^X_{\ell m}(z_i) n^{Y^*}_{\ell' m'}(z_j) \rangle = 0.
\end{equation}
Finally, the angular power spectrum $C_{\ell}^{XY}(z_i,z_j)$ for different tracers and different redshift bins can be written as
\begin{equation}
C_{\ell}^{XY}(z_i,z_j) = \frac{2}{\pi} \int\frac{dk}{k} \mathcal{P}(k) \Delta^{X,z_i}_{\ell}(k) \Delta^{Y,z_j}_{\ell}(k),
\label{eq:Cls}
\end{equation}
where $\mathcal{P}(k)= k^3P(k)$ is the primordial  power spectrum and
\begin{equation}
\Delta^{X,z_i}_{\ell}(k) = \int_{z_i-\Delta z}^{z_i+\Delta z} dz \frac{dN_X}{dz}W(z,z_i)\Delta^X_\ell(k,z),
\label{eq:Delta_l}
\end{equation}
where $\dfrac{dN_X}{dz}$ is the source number density per redshift interval, $W(z,z_i)$ is a window function centered at $z_i$ with half-width $\Delta z$ (with the integral of $\displaystyle W(z,z_i)\frac{dN_X}{dz}$ being normalized to unity). Note that equation \eqref{eq:Cls} follows the notation of reference~\cite{bonvin:cl}, which reflects how the public code \texttt{CLASS}~\cite{blas:class, didio:classgal} is built. The $\Delta^X_\ell(k,z)$ is the angular number count fluctuation of the $X$ tracer, which is determined by density ($\mathrm{den}$), velocity ($\mathrm{vel}$), lensing ($\mathrm{len}$) and gravity ($\mathrm{gr}$) effects~\cite{bonvin:cl, challinor:deltag}:
\begin{equation}
\Delta_\ell(k,z) = \Delta^\mathrm{den}_{\ell}(k,z) + \Delta^\mathrm{vel}_\ell(k,z) + \Delta^\mathrm{len}_\ell(k,z) + \Delta^\mathrm{gr}_\ell(k,z).
\label{eq:numbercount_fluctuation}
\end{equation}
The reader interested in the full expressions of the number counts fluctuations in equation \eqref{eq:numbercount_fluctuation} can find them in Appendix \ref{app:deltas}. The relative importance between each of these terms depends on the specific configuration (redshift bins, window functions, etc.) but some general statements can be made (see e.g., figure 1 of reference \cite{didio:classgal}). The main contribution is usually given by the density term, whereas the gravity effects are subdominant (even by two orders of magnitude at $\ell \sim 50$), since it is relevant mostly at horizon scales. The lensing term is only slightly scale-dependent, while the velocity one can be comparable to it at smaller scales, while stronger than it at lower multipoles (even almost one order of magnitude at very low $\ell$). This last statement holds especially for auto-bin correlations, while for power spectra among distant redshift bins the lensing term can overcome the velocity.

The angular power spectra were computed using \texttt{Multi\_CLASS}, the modified version of \texttt{CLASS} presented in~\cite{bellomo:multiclass, bernal:multiclass} which allows the user to compute cross-correlations between different tracers $X\neq Y$. We consider Gaussian window functions and fix the cosmological parameters from the Planck 2015 results with $n_s=0.9619$, $ln10^{10}A_s=3.0980$, $\Omega_{\mathrm{cdm}}=0.26377$, $\Omega_{\mathrm{b}}=0.04828$, $h = 0.67556$~\cite{Planck:XIII}.

As can be seen by the above equations and Appendix \ref{app:deltas}, there are four main ingredients that are needed to fully characterize one tracer $X$:
\begin{itemize}
    \item \textit{Redshift distribution} $\dfrac{dN_X}{dz}$: the source number density per redshift interval is characterized by a shape which is a fundamental ingredient in the angular power spectra computation (see equation \eqref{eq:Delta_l}). Eventually, the number of sources in a specific redshift bin is also necessary to compute the shot noise that enters in the estimate of the observed $\tilde{C}_\ell$'s in equation \eqref{eq:tilde_Cls}. More details about the redshift distributions of the considered tracers are given in sections \ref{sec:SFR} and \ref{sec:merging_rates}.
    \item \textit{Bias} $b_X$: it quantifies the mismatch between the distribution of matter and of the tracer $X$ (see e.g.,~\cite{Kaiser:bias, Bardeen:bias, Mo:smallhalosbias, matarrese:clusteringevolution, dekel:stochasticbiasing, benson:galaxybias, peacock:halooccupation, Desjacques:bias}). In this work we make use of the linear bias formulation. Indicating the local contrasts of matter and tracer $X$ at position $x$ respectively by $\delta(x)$ and $\delta_X(x)$, we write $\delta_{X}(x) \equiv \frac{n_{X}(x)-\bar{n}_{X}}{\bar{n}_{X}}=b_{X} \delta(x)$, where $n_{X}$ is the comoving density of tracer X and $\bar{n}_{X}$ is its mean value. The bias appears as a linear factor in the density term of equation \eqref{eq:numbercount_fluctuation} (see the full expression in appendix \ref{app:deltas}).
    More details about the bias of our tracers are given in sections \ref{sec:g_bias} and \ref{sec:gw_bias}.
    \item \textit{Magnification bias} $s_X(z)$: it quantifies the change in the observed surface density of sources of tracer X induced by gravitational lensing~\cite{turner:magnificationbias}. Two effects compete against each other: on one side the number of observed sources can increase due to a magnification of the received flux, which would make visible some sources right below the visibility threshold (luminosity or magnitude for galaxies and Signal-to-Noise ratio for GWs); on the other side an increase of the area reduces the observed number density of objects. The magnification bias mainly affects the lensing term of equation \eqref{eq:numbercount_fluctuation}, but enters also in the velocity and gravity terms. Further discussion about magnification bias of our tracers can be found in sections \ref{sec:g_magbias} and \ref{sec:gw_magbias}.
    \item \textit{Evolution bias} $f_X^{\mathrm{evo}}$: it reflects the fact that the number of elements of a tracer X is not necessarily conserved in redshift due to the possible formation of new objects. The evolution bias can be written as~\cite{challinor:evolutionbias, jeong:evolutionbias, bertacca:evolutionbias}:  $f^\mathrm{evo}_\mathrm{g}(z) = \frac{d\ln\left(a^3\frac{d^2N_\mathrm{g}}{dzd\Omega}\right)}{d\ln a}$, where $a$ is the scale factor and $\frac{d^2N_\mathrm{X}}{dzd\Omega}$ is the absolute distribution of objects of tracer $X$, which can usually be substituted by the observed distribution with good approximation~\cite{Scelfo_2018}. The evolution bias appears only in sub-leading contributions, since it is just present in the non-dominant part of the velocity term and in the gravity term (which has a smaller influence with respect to the others) \cite{didio:classgal} of equation \eqref{eq:numbercount_fluctuation}.
\end{itemize}

We complete this section by explicitly summarizing the dependence of the various contributions to the angular number count fluctuations on the three biases types presented above (see appendix \ref{app:deltas} for full expressions):
\begin{equation}
\begin{cases}
\Delta_\ell^\mathrm{den} = \Delta_\ell^\mathrm{den}(b_X) \\
\Delta_\ell^\mathrm{vel} = \Delta_\ell^\mathrm{vel}(s_X, f_X^{\mathrm{evo}}) \\ 
\Delta_\ell^\mathrm{len} = \Delta_\ell^\mathrm{len}(s_X) \\
\Delta_\ell^\mathrm{gr} = \Delta_\ell^\mathrm{gr}(s_X, f_X^{\mathrm{evo}})
\end{cases} 
\end{equation}
where dependencies on $k$ and $z$ are implied.

\section{Tracers}\label{section:tracers}

In this section we describe how our two tracers (galaxies and GWs) are characterized, along with the theoretical frameworks used in their modeling. Since we deal with a statistical approach to count and describe the properties of galaxies and GWs, we use the notation $dp/d\Theta$ to express the probability distribution of a generic variable $\Theta$.

\subsection{Galaxies}

Our first tracers are actively star-forming galaxies. We do not deal with any specific galaxy catalog: we count and distribute galaxies on the basis of the observationally determined SFRF at different redshifts (described in section \ref{sec:SFR}). The SFR of a galaxy measures the stellar mass in solar units formed per year inside the galaxy. We briefly explain how it is measured in section \ref{sec:SFR}. In this work we consider objects with three different SFR lower limits: $\psi\geq 10,\,100,\,300\,  M_\odot /$yr. The lower value of $10 M_\odot /$yr roughly corresponds to the limit below which uncertainties in the star formation rate functions (SFRFs) are significant, especially at high redshift. The cut of $100 M_\odot /$yr is set to take into account the highly star forming dusty galaxies which constitute the bulk of the cosmic SFR. Finally, the highest limit of $300 M_\odot /$yr is set to take into account the most extreme star forming objects.

\subsubsection{SFR functions}
\label{sec:SFR}

The star formation rate functions SFRF = $  d^2N/d\log_{10}\psi/ dV$ correspond to the number density of galaxies per cosmological comoving volume per logarithmic bin of SFR at a given cosmic time $t$ or redshift $z$. In the last years several observations (e.g., UV+far-IR/submillimeter/radio luminosity functions and stellar/gas/dust mass functions) have allowed to robustly estimate these functions. The SFR of a galaxy could in principle be estimated by its UV luminosity, since it is proportional to the quantity of young stars present in the galaxy. However, this estimation can be easily biased by the presence of dust. Indeed, even a modest amount of dust can significantly absorb the UV radiation and re-emit it in the far-IR/(sub)millimeter wavelengths. Standard UV slope corrections (e.g.,~\cite{meurer+99, calzetti+00, bouwens+15}) can still be applied to galaxies with relatively low SFR $\psi\lesssim 30-50 M_\odot/  yr$, since the dust attenuation for them is mild. Therefore, deep UV surveys in the rest frame UV band are enough to robustly determine the SFRF at the faint end. Instead, in highly star-forming galaxies with SFR $\psi\gtrsim 30-50 M_\odot/$yr dust obscuration is heavy and the corrections mentioned above are no more reliable (e.g.,~\cite{silva+98,efstathiou+00,coppin+15,reddy+15,fudamoto+17}). To soundly estimate their SFRF, it is necessary to use far-IR/(sub)millimeters observations. The latters have been exploited in many works over the recent years (e.g.,~\cite{lapi+11,gruppioni+13,gruppioni+15,gruppioni+19,magnelli+13}) to reconstruct, in combination with UV data, the SFRF for the whole SFR range at redshift $z\lesssim 3$. At higher redshifts, given the sensitivity limits of wide-area far-IR surveys, the reconstruction of the SFRF, especially at the bright end, is more uncertain. Useful information have been obtained from far-IR/(sub)millimeter stacking (see~\cite{rowan+16},~\cite{dunlop+17}) and super-deblending techniques (see~\cite{liu+18}), from targeted far-IR/(sub)millimeter observations (e.g.,~\cite{riechers+17,marrone+18,zavala+18}) and from radio surveys (\cite{novak+17}).

All the above datasets have been fitted through simple Schechter functions by~\cite{mancuso+16}, obtaining:
\begin{equation}
    \frac{  d^2N}{  d\log_{10}\psi\, dV}(\log_{10}\psi,z)=\mathcal{N}(z)\left(\frac{\psi}{\psi_{  c}(z)}\right)^{1-\alpha(z)}  e^{-\psi/\psi_{  c}(z)}\, ,
\end{equation}
where the values of the redshift-dependent parameters $\mathcal{N}(z)$, $\psi_{  c}(z)$ and $\alpha(z)$ can be found in table 1 of~\cite{mancuso+16} (see also figure 1 of~\cite{boco+19}). From the SFRF we can obtain the number density of galaxies per unit comoving volume at different cosmic times $t$ and the cosmic star formation rate density as:
\begin{align}
    \frac{  dN}{  dV}(t)&=\int  d\log_{10}\psi\,\frac{  d^2N}{  d\log_{10}\psi\, dV}\,(\log_{10}\psi, t)\label{eq|number density}\, ,\\
    \rho_\psi(t)&=\int  d\log_{10}\psi\;\psi\,\frac{  d^2N}{  d\log_{10}\psi\,dV}\,(\log_{10}\psi, t)\, .
\end{align}

Notice that the evolution of the cosmic star formation rate density with redshift, reconstructed in this way, is well in agreement with the available datasets (see~\cite{boco+19}). The number of galaxies per redshift bin $dN/dz$ can be easily obtained multiplying equation \ref{eq|number density} by the differential cosmological comoving volume $  dV/dz$. The redshift distribution of our 3 galactic populations (galaxies with $\psi>10,\,100,\,100\,M_\odot/$yr) can be obtained integrating the SFRF excluding galaxies below a certain threshold $\bar{\psi}$:
\begin{equation}
    \frac{  dN_{\bar\psi}}{  dz}(t,\psi\geq\bar\psi)=\frac{  dV}{  dz}\int_{\bar\psi}  d\log_{10}\psi\frac{  d^2N}{  d\log_{10}\psi \, dV}.
    \label{eq|dndz bin}
\end{equation}

In figure \ref{fig:specifics_g}, left panel, the three $  dN_{\bar\psi}/dz$ of the galactic populations considered in this work are plotted as a function of redshift.

From this quantity it is straightforward to compute the evolution bias for each galactic population as
\begin{equation}
f^\mathrm{evo}_\mathrm{g}(z) = \frac{d\ln\left(a^3\frac{d^2N_{\bar\psi}(z,\psi\geq\bar\psi)}{dzd\Omega}\right)}{d\ln a}.    
\end{equation}

\subsubsection{Galaxy bias}
\label{sec:g_bias}
A fundamental ingredient entering in the computation of the observed number counts fluctuations in equation \ref{eq:numbercount_fluctuation} is the bias $b_X(z)$. Since our tracers are galaxies selected, counted and divided by their SFR, we should connect the bias to this quantity. We adopt the procedure of reference~\cite{aversa+15} associating the luminosity/SFR of the galaxy to the mass of the hosting dark matter halo through an abundance matching technique and then assigning to a galaxy with given SFR the bias of the corresponding halo. Abundance matching is a standard method to derive a monotonic relationship between the galaxy and the halo properties by matching the corresponding number densities in the following way:
\begin{equation}
    \int_{\log_{10}\psi}^\infty  d\log_{10}\psi'\frac{  d^2N}{  d\log_{10}\psi'\,dV}=\int_{-\infty}^\infty  d\log_{10} M_H'\frac{  d^2N}{  d\log_{10} M_H'\,dV}\frac{1}{2}  erfc\left\{\frac{\log_{10}{(M_H(\psi)/M_H')}}{\sqrt{2}\tilde{\sigma}}\right\}\, ,
    \label{eq:abundance}
\end{equation}
where $M_H'$ is the halo mass, $  d^2N/d\log_{10} M_H'/dV$ is the galaxy halo mass function i.e. the mass function of halos hosting one individual galaxy (see Appendix A of~\cite{aversa+15}) and $M_H(\psi)$ is the relation we are looking for. Finally, $\tilde{\sigma}\equiv\sigma\,  d\log_{10} M_H/d\log_{10}\psi$ is the scatter around that relation (we set $\sigma_{\log_{10}\psi}\simeq 0.15$ following~\cite{lapi+2006,lapi+2011,lapi+2014}). Once $M_H(\psi)$ is determined we assign to each galaxy the bias corresponding to the halo associated to its SFR: $  b(z,\psi)=b(z,M_H(z,\psi))$, where $b(z,M_H)$ is computed as in~\cite{sheth+01} and approximated by~\cite{lapi+14}. Note that this formulation for the bias is based on the excursion set approach. Other possible alternatives are present, such as the Effective Field Theory, Peak Theory, etc. (see e.g., \cite{Desjacques:bias, Bernardeau:bias} and references therein).

It is now easy to compute an effective bias for all the galaxies above a certain SFR threshold $\bar{\psi}$ weighting $b(z,\psi)$ by the corresponding galaxy distribution:
\begin{equation}
      b_{\bar\psi}(z,\psi\geq\bar\psi)=\frac{\int_{\log_{10}\bar\psi}^\infty  d\log_{10}\psi\frac{  d^2N}{  d\log_{10}\psi\,dV}b(z,\psi)}{\int_{\log_{10}\bar\psi}^\infty  d\log_{10}\psi\frac{  d^2N}{  d\log_{10}\psi\,dV}}\, .
\end{equation}

In figure \ref{fig:specifics_g}, middle panel, we show the galaxy effective bias as a function of redshift for our galactic populations. We see, as expected, that it tends to increase with redshift and it is, in general, lower for galaxies with lower SFRs, since these are typically associated to less massive halos.

Note that the bias computed in this section refers to star forming galaxies, which can be the progenitors of quenched massive objects at the present time. We stress that, in order to estimate the bias for these galaxies, we should not look at their current SFR, but at the SFR of their progenitor when most of the stellar mass was accumulated, which is directly linked to the mass of the host dark matter halo, via the abundance matching technique described in equation \eqref{eq:abundance}.

\subsubsection{Galaxy magnification bias}
\label{sec:g_magbias}

The magnification bias is another important factor entering in the angular power spectra computation. In fact, as we already described in section \ref{sec:methods}, the contribution to the angular number count fluctuations due to the lensing term of equation \eqref{eq:numbercount_fluctuation} can be comparable to that of the velocity term (or even higher, especially when correlating objects between distant redshift bins). As mentioned above, our galactic populations have SFR cuts $\bar\psi=10,\,100,\,300\,M_\odot/$yr. The magnification bias for each of them is proportional to the logarithmic slope of their $  dN_{\psi}/dz$ computed at $\psi=\bar\psi$:
\begin{equation}
      s_{  g,\bar\psi}(z)=-\frac{2}{5}\left.\frac{  d\log_{10}{\left(\frac{  d^2N_{\psi}(z,>\psi)}{  dzd\Omega}\right)}}{  d\log_{10}\psi}\right|_{\psi=\bar\psi} \, .
\end{equation}
Using equation (\ref{eq|dndz bin}) we can show that the magnification bias can be directly related to the SFRF as:
\begin{equation}
      s_{  g,\bar\psi}(z)=\frac{2}{5\ln 10}\frac{\frac{  d^2N}{  d\log_{10}\psi\,dV}(z,\bar\psi)}{  dN_{\bar\psi}/  dz}\frac{  dV}{  dz}\, .
\end{equation}

\begin{figure}[ht!]
	\centerline{
		\subfloat{
			\includegraphics[width=1.1\linewidth]{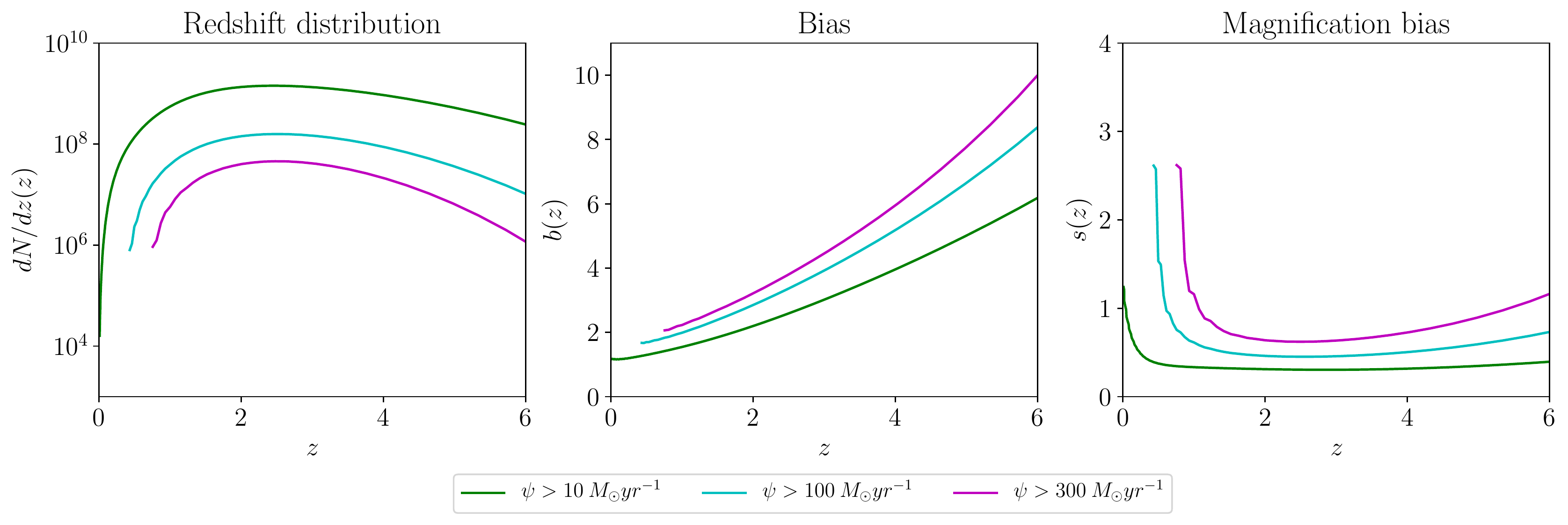}}}
	\caption{Full-sky redshift distributions (\textit{left}), bias (\textit{center}) and magnification bias (\textit{right}) for all galactic populations. Quantities referred to galaxies with Star Formation Rate $\psi > 10 M_\odot/yr, \: \psi > 100 M_\odot/yr, \: \psi > 300 M_\odot/$yr are respectively in green, cyan and magenta lines.}
	\label{fig:specifics_g}
\end{figure}

Figure \ref{fig:specifics_g} (right panel) shows the magnification bias for our galactic populations as a function of redshift. We can see that, at small $z$, the magnification bias decreases rapidly, especially for the tracers with higher SFRs: this is due to the fact that at small redshifts we have less and less galaxies with high SFRs, therefore the function $  d^2N(z,\psi\geq\bar\psi)/dz/d\Omega$ strongly depends on the choice of the faint end in SFR. Moreover, the overall magnification bias for higher star forming galaxies is larger because they are less and a variation of the faint end SFR limit has a larger impact on their $  dN/dz$. This is why, in general, the magnification bias shape tends to be specular to the one of the $  dN/dz$.

\subsection{Gravitational waves}
\label{sec:tracer_GW}

The other class of tracers that we are considering are GWs resolved signals coming from the merging of BH-BH, NS-NS and BH-NS binaries. In the next subsections, as already done for the case of galaxies, we are going to illustrate how to compute their merging and detected rates, their bias and magnification bias.

\subsubsection{Merging rates}
\label{sec:merging_rates}

The merging rate of compact binaries is the number of merging per year per redshift bin $  d\dot{N}/dz$\footnote{The factor $  d\dot{N}/dz$ is well defined since it stays put for typical observation times of the order of $\sim 10\,\rm yr$.}. Multiplying it by an observational time one gets the number of events per redshift bin in that period of time. Computing the merging and detected rates of compact binaries is a complex issue since it involves the necessity to understand and to correctly model many astrophysical processes occurring on different time and spatial scales: from stellar astrophysics, to galaxy formation and evolution, to GW physics. However, a number of studies have approached the problem combining population synthesis simulations (e.g.,~\cite{dominik+13,dominik+15,demink+13,spera+15,spera+17,spera+19,giacobbo+18}) either with cosmological simulations (e.g.,~\cite{mapelli+17,oshaughnessy+17,lamberts+17,mapelli+18,artale+19}) or with recipes on the cosmic star formation rate density and metallicity distributions inferred from observations (e.g.,~\cite{belczynski:massivebhsmergers,lamberts+16,cao+18,elbert+18,li+18,boco+19,neijssel+19}). In this work we follow the approach of reference~\cite{boco+19}, briefly sketched hereafter.

The three main ingredients to compute the merging rates of compact binaries are: \textit{i)} an observational determination of the SFRF at different redshifts, \textit{ii)} average chemical enrichment histories of individual galaxies and \textit{iii)} outcomes from single stellar and binary evolution simulations.

The first ingredient has already been described in section \ref{sec:SFR} and provides the galaxy statistics. In the following subsections we are going to describe respectively the other two ingredients and the way to combine them to compute the merging rates of compact binaries.

\subsubsection*{Metallicity}\label{subsection:metallicity}

The average chemical enrichment histories of individual galaxies is crucial since it allows to associate a metallicity Z to galaxies with different properties (SFR, mass, age or morphological type). Knowing the metallicity is fundamental because many binary evolutionary phenomena strongly depend on it: stellar winds, supernova kicks, direct collapse, common envelope effects, etc. (for a more detail explanation of the main effects of metallicity on stellar and binary evolution see section \ref{sec:Z_dependency} and references therein). In~\cite{boco+19} the chemical enrichment history of a galaxy with a given average SFR is reproduced with a simple model featuring a linear increase of the metallicity in the early stages of the galaxy life up to a saturation value dependent on the SFR. This model is an approximation of the more elaborated chemical evolution model of~\cite{pantoni+19} and~\cite{lapi+20} and well reproduces observations of both elliptical and disk galaxies (see e.g.,~\cite{arrigoni+10,spolaor+10,gallazzi+14,andrews+13,zahid+14,delarosa+16,onodera+16}). The authors of~\cite{boco+19} generate a metallicity distribution at given cosmic time $t$ and SFR $  dp/d\log_{10} Z\,(\log_{10} Z|t,\psi)$ taking into account the time spent by a galaxy with given SFR in each bin of metallicity. Their final expression to compute the metallicity distribution is:
\begin{equation}
    \frac{  dp}{  d\log_{10} Z}(\log_{10} Z|t,\psi)=\Delta\times\frac{Z}{Z_{  sat}}\ln(10)\Theta_H(Z-Z_{  sat})+(1-\Delta)\times\delta_D(\log_{10} Z-\log_{10} Z_{  sat})\, ,
\end{equation} 
where $Z_{  sat}(t,\psi)$ and $\Delta(t,\psi)$ are parameters depending on the cosmic time and the SFR of the single galaxy. $Z_{  sat}$ represents the saturation value of the metallicity and its typical values are in the range $\sim 0.3-1.5 Z_\odot$, $\Delta\sim 0.1-0.3$ specifies how quickly the metallicity saturates to such values as a consequence of the interplay between cooling, dilution and feedback processes. 

\subsubsection*{Stellar and binary evolution}

The outcomes of single stellar and binary evolution simulations can provide three important factors: the remnant mass $m_\bullet(m_\star, Z)$ as a function of the zero age main sequence (ZAMS) star mass $m_\star$ and metallicity $Z$, a time delay distribution between the formation of the binary and the merging $  dp/dt_{  d}$ and a mass ratio distribution $  dp/dq$, where $  q$ is the ratio between the less and the more massive compact remnant.

For the mass distribution of compact remnants we take as a reference the $m_\bullet(m_\star,Z)$ relation given in~\cite{spera+17}, and we generate a probability distribution function just applying a logarithmic gaussian scatter ($\sigma=0.1  $dex) around the $m_\bullet(m_\star,Z)$ value to take into account possible uncertainties coming from stellar evolutionary processes as in~\cite{boco+19}:
\begin{equation}
    \frac{  dp}{  d\log_{10} m_\bullet}(m_\bullet|m_\star,Z)=\frac{1}{\sqrt{2\pi}\sigma }\exp{\left[-(\log_{10} m_\bullet-\log_{10} m_\bullet(m_\star,Z))^2/2\sigma^2\right]}\, .
\end{equation}
To get rid of the dependence on the initial stellar mass it is sufficient to select an initial mass function (IMF) $\phi(m_\star)$ (in this work we used the Chabrier one~\cite{chabrier+03}\footnote{In principle the IMF could be different and also dependent on the galaxy properties as SFR or metallicity. However the authors of reference~\cite{boco+19} showed that its choice has only a mild impact on the detected GWs rate.}) and integrate over the initial stellar masses weighting the integral with the IMF:
\begin{equation}
    \frac{  dp}{  d\log_{10} m_\bullet}(m_\bullet|Z)=\int_{\bar{m}_\star}  dm_\star\phi(m_\star)\frac{  dp}{d\log_{10} m_\bullet}(m_\bullet|m_\star,Z)\, ,
    \label{eq|remnant mass pdf}
\end{equation}
where $\bar{m}_\star\sim 7M_\odot$ is the ZAMS star mass limit originating a NS remnant\footnote{Note that in equation \eqref{eq|remnant mass pdf}, the integral should contain the quantity $\phi(m_\star)/\int dm_\star\phi(m_\star)m_\star$. However, in literature the denominator is usually left implicit because of the IMF normalization condition $\int dm_\star\phi(m_\star)m_\star=1M_\odot$, though the reader should keep track of the measurement units (i.e. the factor computed in equation \eqref{eq|remnant mass pdf} is a distribution of remnant masses per unit of star formed mass).}. However, since the amplitude of the gravitational wave events is determined by the chirp mass $\mathcal{M}_{\bullet\bullet}\equiv m_\bullet  q^{3/5}/(1+q)^{1/5}$, rather than by the primary mass $m_\bullet$ \footnote{The primary mass $m_\bullet$ is the mass of the most massive compact remnant of the binary.}, we should make use of the mass ratio distribution to change variable and determine the probability distribution function for a given chirp mass in the following way:
\begin{equation}
    \frac{  dp}{  d\mathcal{M}_{\bullet\bullet}}(\mathcal{M}_{\bullet\bullet}|Z)=\int  dq\frac{  dp}{  dq}\frac{  dp}{  dm_\bullet}(m_\bullet(\mathcal{M}_{\bullet\bullet},q)|Z)\frac{  dm_\bullet}{  d\mathcal{M}_{\bullet\bullet}}(  q)\, ,
    \label{eq|chirp mass distribution}
\end{equation}
where $m_\bullet(\mathcal{M}_{\bullet\bullet},q)=\mathcal{M}_{\bullet\bullet}(1+  q)^{1/5}/q^{3/5}$, $dm_\bullet/ d\mathcal{M}_{\bullet\bullet}=(1+  q)^{1/5}/q^{3/5}$ and the distribution $  dp/dq$ is taken from binary evolution simulations. In particular, the mass ratio distribution for BH-BH mergers scales linearly with $  q$: $  dp/dq\propto q$ (see e.g.,~\cite{demink+13, belczynski:massivebhsmergers, kovetz:bhbh}), instead, for NS-NS or BH-NS merging, the mass ratio distribution tends to be flatter (see~\cite{dominik+12,dominik+15,demink+15,chruslinska+19a,mapelli+18}). We set the $q$ distributions ranges for the three types of merging events on the basis of the allowed masses for each CO type. The authors of reference~\cite{boco+19} have checked that the merging rate depends very little on the chosen $q$ distribution.

The last ingredient provided by observations and simulations is the probability distribution function for the time delay between the formation of the binary and its merging: $  dp/dt_d\propto t_d^{-1}$ ~\cite{dominik+12, giacobbo+18}, normalized to unity between a minimum value of $  t_{d,min}\sim 50$Myr and the age of the universe.

\subsubsection*{Computing merging and detected rates}\label{section:merging_rates}

After this brief overview of the main ingredients, we can compute the cosmic merging rates of compact remnant binaries per redshift and chirp mass interval in the following way:
\begin{equation}
\begin{split}
    \frac{  d^2\dot{N}_{  merge}}{  dz\,d\mathcal{M}_{\bullet\bullet}}(t,\mathcal{M}_{\bullet\bullet})=&f_{\rm  eff}\frac{  dV}{  dz\,(1+z)}\int  dt_{  d}\frac{  dp}{  dt_{  d}}\int  d\log_{10}\psi\frac{  d^2N(\log_{10}\psi, t-t_{  d})}{  d\log_{10}\psi\,dV}\psi\times\\
    &\times\int  d\log_{10} Z\frac{  dp}{  d\log_{10} Z}(\log_{10} Z|t-t_{  d},\psi)\frac{  dp}{  d\mathcal{M}_{\bullet\bullet}}(\mathcal{M}_{\bullet\bullet}|Z)\, .
\end{split}
\label{eq|rmerge}
\end{equation}

To grasp the meaning of this cumbersome expression let us look first at the innermost integral: it represents the probability of formation of a compact remnant with chirp mass $\mathcal{M}_{\bullet\bullet}$ in a galaxy with SFR $\psi$. This quantity is then integrated over all the galaxies weighted with the galaxy statistics (the SFRF). Since we are computing the merging rates at the cosmic time $t$, the quantities related to the formation of the binary should be computed at $t-t_d$, with $t_d$ being the time delay between the formation and the merging of the binary. The outermost integral over the time delay is finally performed to account for all the time delays. The factor $  dV/dz$ is just the differential comoving volume, while the $(1+z)$ factor in the denominator keeps into account the cosmological time dilation. Finally, the factor $f_{\rm  eff}$ is defined as the fraction of primary compact remnants hosted in binary systems with characteristics apt to allow merging within a Hubble time. 

The factor $f_{\rm eff}$ is the result of many different and complex physical processes related to stellar and dynamical evolution (binary fraction, common envelope development/survival, SN kicks, mass transfers, etc.), so it could in principle depend on metallicity and binary type. In stellar and binary evolution simulations (e.g.,~\cite{oshaughnessy+10,dominik+15,belczynski:massivebhsmergers,spera+17,giacobbo+18,chruslinska+18,neijssel+19}) this quantity is naturally obtained, but largely dependent on model assumptions. This is why, at this stage, as already done in other works (e.g. ~\cite{cao+18},~\cite{li+18},~\cite{Calore:crosscorrelating}), we set it empirically by normalizing the local BH-BH, BH-NS and NS-NS merger rates to the logarithmic average values of the $90\%$ confidence interval measured by the LIGO/Virgo collaboration after the O1 and O2 runs (see~\cite{abbott+19}): $30\,\rm Gpc^{-3}/$yr for BH-BH, $650\,\rm  Gpc^{-3}/$yr for NS-NS and $25\,\rm  Gpc^{-3}/$yr for BH-NS (this last choice is less certain since the BH-NS local merging rate is limited only by an upper value). We stress that in doing so the factor $f_{\rm  eff}$ loses all the possible metallicity dependence, which, however, is highly uncertain and model dependent (we will come back to its metallicity behaviour in section \ref{sec:discerning_scenarios}). Therefore, this factor acts only on the normalization of the merging rates and can be changed when a more accurate determination of the local rates will be done after further GWs observations. Thus, the difference in the merging rates normalization between the three types of merging events is given by the different $f_{\rm  eff}$ factors, while the difference in shape is given by the distribution $  dp/d\mathcal{M}_{\bullet\bullet}(\mathcal{M}_{\bullet\bullet}|Z)$ in equation (\ref{eq|chirp mass distribution}), depending on the stellar and binary evolution prescriptions.

Once the merging rates per chirp mass bin are computed, it is easy to derive the detected rates by a specific GW detector. As mentioned in section \ref{sec:intro}, we consider the ET instrument. The rates per unit redshift, chirp mass and signal to noise ratio (SNR) can be computed as:
\begin{equation}
    \frac{  d^3\dot{N}_{merge}}{  dz\,d\mathcal{M}_{\bullet\bullet}\,d\rho}(\rho|z,\mathcal{M}_{\bullet\bullet})=\frac{  d^2\dot{N}_{  merge}}{  dz\,d\mathcal{M}_{\bullet\bullet}}\frac{  dp}{  d\rho}(\rho|z,\mathcal{M}_{\bullet\bullet})
    \label{eq|mergingrates_rho}
\end{equation}
where $  dp/d\rho$ is the probability distribution of SNR dependent on redshift, chirp mass and on the sensitivity curve of the detector (for a full treatment of the $  dp/d\rho$ see~\cite{taylor+12,li+18,boco+19,boco+20}). Therefore, the rates with SNR $\rho>\bar\rho$ can be computed as:
\begin{equation}
    \frac{  d\dot{N}_{\bar\rho}}{  dz}(z,\rho\geq\bar\rho)=\int_{\bar\rho}  d\rho\frac{  d^2\dot{N}}{ 
    dz\,d\rho}(\rho,z)=\int_{\bar\rho}  d\rho\int  d\mathcal{M}_{\bullet\bullet}\frac{  d^3\dot{N}}{  dz\,d\mathcal{M}_{\bullet\bullet}\,d\rho}(\mathcal{M}_{\bullet\bullet}|z,\rho) \, .
\end{equation}

We consider the GW event detected when the SNR is higher than $\bar\rho=8$. The detected rates by ET for BH-BH, NS-NS and BH-NS are shown in figure \ref{fig:specifics_gw} (left panel).

Eventually, the rate for GWs events above a given SNR can be used to compute the evolution bias:
\begin{equation}
f^\mathrm{evo}_\mathrm{GW}(z) = \frac{d\ln\left(a^3\frac{d^2\dot{N}_{\bar\rho}(z,\rho\geq\bar\rho)}{dzd\Omega}\right)}{d\ln a}.    
\end{equation}

\subsubsection{Bias for GW events}
\label{sec:gw_bias}

Since we consider GWs produced by the merging of COs of stellar origin, their signals originate from galaxies and trace their distribution, so that they trace the underlying total matter distribution the same way their host galaxies do. For this reason, GWs events can be characterized by the same bias of their hosts. Since the COs mergers take place in all galaxy types with different rates, the correct estimate of their bias needs to take into account which galaxy types are contributing most/less to the detected mergers, giving proportioned weight to their bias values when estimating that of all GWs events.

In order to assign a redshift dependent bias to the GW events, we make use of the bias $b(z,\psi)$, computed in section \ref{sec:g_bias}, associated to a galaxy at a given redshift with given SFR and we weight it through the quantity $  d^3\dot{N}_{merge}/dz/d\rho/d\log_{10}\psi$ which keeps into account the contribution of the different SFRs (i.e. of different galaxies) to the 
total merging rates at a given redshift and SNR. This differential merging rate can be computed from equation \eqref{eq|rmerge} and \eqref{eq|mergingrates_rho} not integrating over the SFR. Therefore, to compute the bias for gravitational waves we use the following expression:
\begin{equation}
    b_{  GW}(z,\rho)=\frac{\int  d\log_{10}\psi\frac{  d^3\dot{N}}{  dz\,d\rho\,d\log_{10}\psi}b(z,\psi)}{\int  d\log_{10}\psi\frac{  d^3\dot{N}}{  dz\,d\rho\,d\log_{10}\psi}}\, .
\end{equation}

The effective bias, i.e. the bias for GWs with a SNR above a certain threshold $\bar{\rho}$, is now easy to compute:
\begin{equation}
    b_{  GW,\bar{\rho}}(z,>\bar{\rho})=\frac{\int_{\bar{\rho}}  d\rho\frac{  d^2\dot{N}}{  dz\,d\rho}b(z,\rho)}{\int_{\bar{\rho}}  d\rho\frac{  d^2\dot{N}}{  dz\,d\rho}}\, .
\end{equation}

The bias for the detected events ($\bar\rho=8$) is shown in figure \ref{fig:specifics_gw} (middle panel). The interpretation of the shape of the GW bias is not trivial and explained in the following. At low redshift its value is $\sim 1$ since the only galaxies that contribute to the GW signals have low SFR and consequently a smaller bias. The following rapid increase with redshift is due to two factors: the first is just the standard growth with redshift of the galaxy bias, the second is that, increasing the redshift, there are more and more highly star forming galaxies that contribute to the GW events. These galaxies, as shown in figure \ref{fig:specifics_g}, are more biased. At redshift $z\sim 5$ the GW bias flattens. Again, this is due to different astrophysical effects. In particular, the redshift increase of the galaxies bias is compensated by the fact that at high redshift the detected GW events receive a larger contribution by less star forming and, thus, less biased galaxies. This is due to two facts: firstly, the number of highly star forming galaxies tends to decrease at redshift $z\gtrsim 3-4$; secondly, in galaxies with high SFR the metallicity is also high and, consequently, the compact remnants produced are less massive. This means that galaxies with larger SFRs tend to produce GW events with lower chirp mass (see section \ref{sec:Z_dependency}). However, at high redshift the detector starts not to see anymore these low chirp mass events and, due to this selection effect, the GW events detected at higher and higher redshifts come from galaxies with lower SFR and are, consequently, less biased.

\subsubsection{Magnification bias for GW events}
\label{sec:gw_magbias}

\begin{figure}[ht!]
	\centerline{
		\subfloat{
			\includegraphics[width=1.1\linewidth]{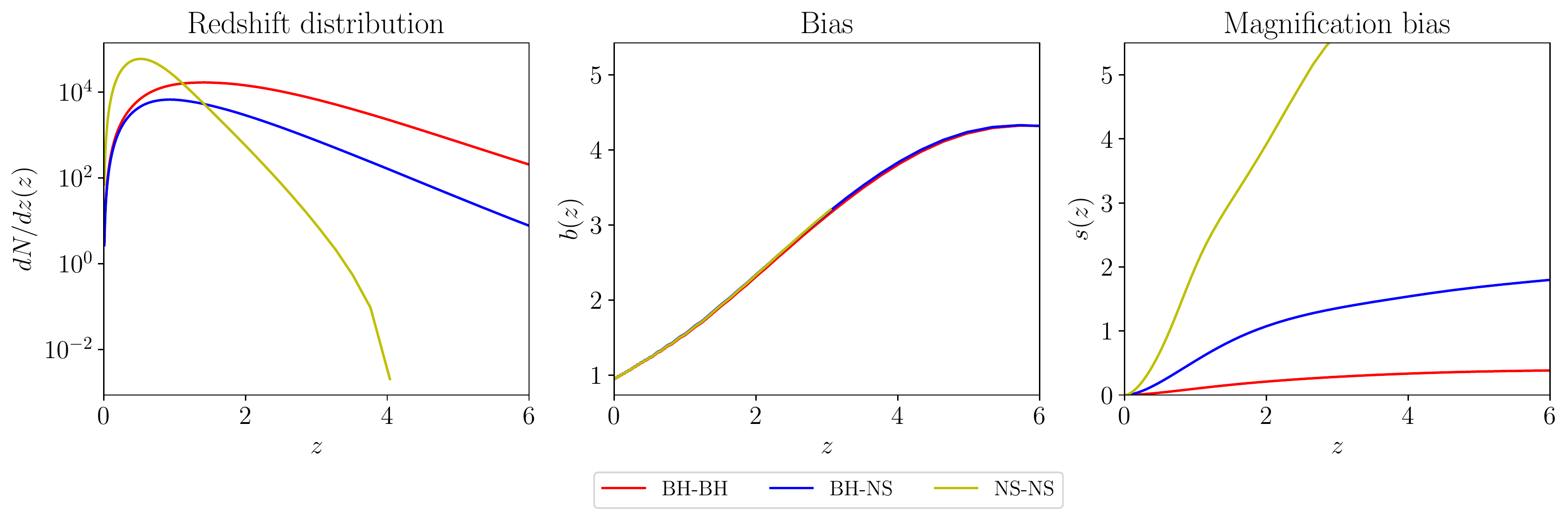}}}
	\caption{Full-sky redshift distributions for an observation time $T_{\mathrm{obs}}=1$yr (\textit{left}), bias (\textit{center}) and magnification bias (\textit{right}) for all GWs tracers, as detected by ET. Quantities referred to BH-BH, BH-NS, NS-NS mergers are respectively in red, blue and yellow lines.}
	\label{fig:specifics_gw}
\end{figure}

\begin{figure}[ht!]
	\centerline{
		\subfloat{
			\includegraphics[width=1.1\linewidth]{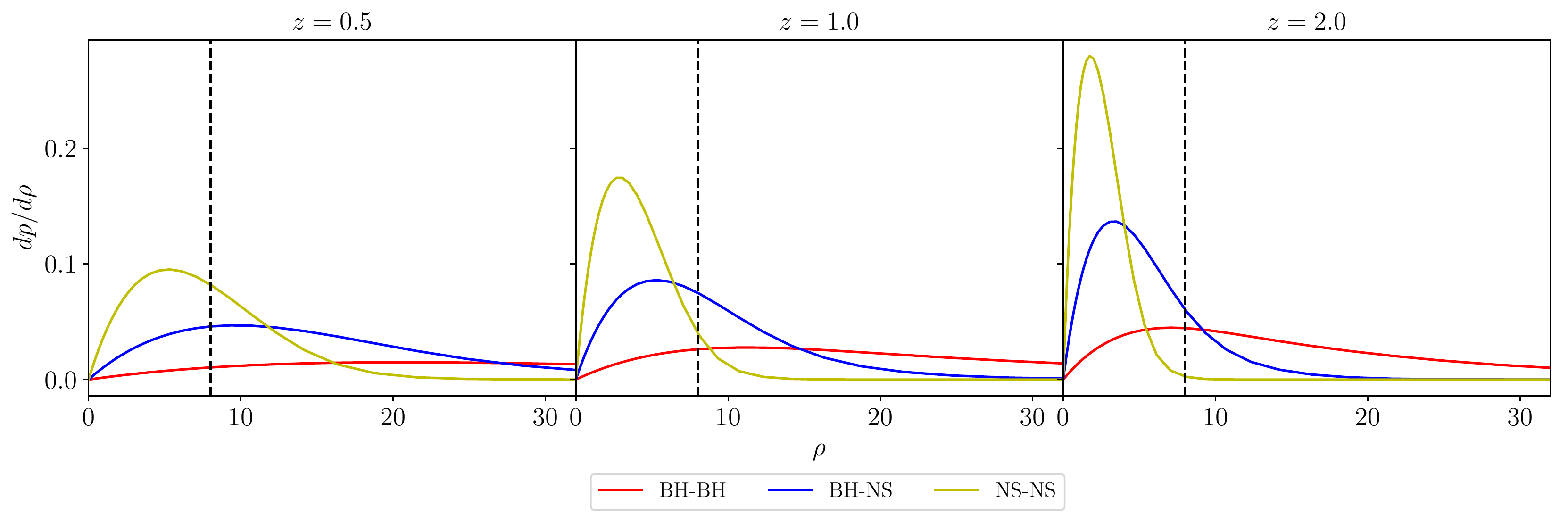}}}
	\caption{SNR $\rho$ normalized distributions at redshifts $z=0.5$ (\textit{left}), $z=1.0$ (\textit{center}) and $z=2.0$ (\textit{right}) for all GWs tracers. Distributions for BH-BH, BH-NS, NS-NS are respectively in red, blue and yellow lines. The black dashed vertical lines correspond to the limit of $\rho = 8$.}
	\label{fig:dpdrho_gw}
\end{figure}

Similarly to the galaxy case and as done in ref.~\cite{Scelfo_2018}, the magnification bias for GW events with $\rho>\bar\rho$ is the logarithmic slope of their $  dN_{\rho}/dz\,(z,>\rho)$ computed at $\rho=\bar\rho$:
\begin{equation}
    s_{  GW,\bar\rho}(z)=-\left.\frac{  d\log_{10}{\left(\frac{  d^2\dot{N}_{\rho}(z,>\rho)}{  dz\,d\Omega}\right)}}{  d\rho}\right|_{\rho=\bar\rho}\, ,
\end{equation}
which, after some algebraic manipulation, can be rewritten as:
\begin{equation}
    s_{  GW,\bar\rho}(z)=\bar\rho\frac{\frac{  d^2\dot{N}(z,\bar\rho)}{  dz\,d\rho}}{  d\dot{N}_{\bar\rho}/dz}\, .
\end{equation}
We shown in figure \ref{fig:specifics_gw} (right panel) the magnification bias for detected mergers ($\bar\rho=8$). It can be seen that the magnification bias for NS-NS events features a fast growth with redshift because the NS-NS distribution in SNR is peaked at lower values of $\rho$ with respect to BH-BH or BH-NS events (see figure \ref{fig:dpdrho_gw}). So, as the redshift increases, the peak of such distribution shifts toward values $\rho\lesssim 8$: the choice of the faint end of SNR has then a huge effect on NS-NS events. Instead, for BH-BH and BH-NS events, the distribution in SNR ratio is much broader, even at high redshifts: the choice of the faint end of SNR has not a large impact on the number of detections. For this reason the magnification bias for those events always remains at moderate values.

In figure \ref{fig:dpdrho_gw} we show the SNR probability distribution functions for BH-BH, BH-NS and NS-NS events at redshift $z=0.5$ left panel, $z=1$ middle panel and $z=2$ right panel.

A useful thing to notice is that changing the cosmological parameters values affects the description of the tracers (both galaxies and GWs) only as a volume term $dV/dz$ in the computation of the redshift distributions, as can be seen in equations \eqref{eq|dndz bin} (for galaxies) and \eqref{eq|rmerge} (for GWs). This implies that, changing the cosmological parameter values, all the redshift distributions will vary in the same way, making it difficult to constrain them.

\section{Galaxies - GWs cross-correlations and Signal-to-Noises}
\label{sec:Cl_SN}
Making use of the formalism presented in section \ref{sec:methods} we compute the cross-correlation angular power spectra $C_\ell$'s between the tracers presented in section \ref{section:tracers}. Note that the GWs events considered in the cross-correlation are computed from the whole galaxy distribution, not just from galaxies with the specified SFR cuts, in order to realistically take into account all the GWs signals that can be detected by a specific instrument (ET in our case). We considered both tomographic and non tomographic approaches, whose details and results are provided in sections \ref{sec:Cl_SN_YESTOMO} and \ref{sec:Cl_SN_NOTOMO} respectively.

Following ref.~\cite{Scelfo_2018}, we can organize the angular power spectra from different tracers and redshift bin couples in a data vector $\mathbf{C}_\ell$ ordered as
\begin{equation}
\mathbf{C}_\ell = 
\begin{pmatrix}
{C}_\ell^{\rm{g}\rm{g}}(z_1,z_1) \\
\vdots	\\
{C}_\ell^{\rm{g}\rm{GW}}(z_1,z_1) \\
\vdots	\\
{C}_\ell^{\rm{GW}\rm{GW}}(z_1,z_1) \\
\vdots
\end{pmatrix}
\label{eq:vectors}
\end{equation}
where $\rm g$ and $\rm GW$ respectively refer to our galaxy and gravitational wave tracers. Given $N_{\mathrm{bin}}$ redshift bins, the $\mathbf{C}_\ell$ is a $N_{\mathrm{bin}}^2$ dimensional data vector, where
its $I^{\mathrm{th}}$ element can be associated to two indices $(I_1,I_2)$, corresponding to the two tracers and redshift bins of the angular power spectra in that specific entry. As an example, the first ($I=1$) entry is associated to the couple of indices $[I_1=\mathrm{\rm g}_{z_1}, I_2=\mathrm{\rm g}_{z_1}]$. From this, one can write the covariance matrix $\left(\mathrm{Cov}_{\ell}\right)_{IJ}$, whose elements are given by
\begin{equation}
\left(\mathrm{Cov}_{\ell}\right)_{IJ} = \tilde{C}^{I_1J_1}_\ell\tilde{C}^{I_2J_2}_\ell+\tilde{C}^{I_1J_2}_\ell\tilde{C}^{I_2J_1}_\ell,
\label{eq:covariance_matrix_chi2}
\end{equation}
where the $\tilde{C_\ell}$ are the angular power spectra of equation \eqref{eq:tilde_Cls}.

In order to characterize the magnitude of the signal that could be extracted by the cross-correlations and determine whether it could be discerned from the noise, it is useful to compute a Signal-to-Noise ratio (S/N). With this purpose, we compute two types of S/N. 

The first one is an estimate of the S/N of the $C_\ell$'s for each combination of redshift bins (at fixed tracers couple). This provides a total of $N_{\mathrm{bin}} \times N_{\mathrm{bin}}$ S/N values. While on the one hand they do not take into account correlations with the other redshift bins (which can be non negligible, especially due to lensing effects), on the other hand they provide an unpacked information about which bin combinations are the most powerful in terms of signal. At fixed multipole $\ell$, the S/N computed in this way can be written as:
\begin{equation}
\biggl(\dfrac{S}{N}\biggl)^2_{[I_1,I_2]}(\ell) = f_{\mathrm{sky}}(2\ell + 1) \dfrac{\bigl(\tilde{C}_\ell^{[I_1,I_2]}\bigl)^2}{\sigma^2_{[I_1,I_2]}(\ell)} = f_{\mathrm{sky}}(2\ell + 1) \dfrac{\bigl(\tilde{C}_\ell^{[I_1,I_2]}\bigl)^2}{\Biggl[\tilde{C}_\ell^{[I_1,I_1]} \tilde{C}_\ell^{[I_2,I_2]} + \bigl(\tilde{C}_\ell^{[I_1,I_2]}\bigl)^2\Biggl]}
\label{eq:SNR_1}
\end{equation}
where $\sigma^2_{[I_1,I_2]}(\ell)$ is obtained imposing $I=J$ in equation \eqref{eq:covariance_matrix_chi2}.

The second method for the computation of the S/N provides (still at fixed tracers couple) one single S/N estimate for the whole probe, taking into account also the covariance between the $C_\ell$'s of different redshift bins. It is computed as:
\begin{equation}
\label{eq:SNR_2}
  \biggl(\dfrac{S}{N}\biggl)^2_{\mathrm{TOT}}(\ell) = f_{\mathrm{sky}}(2\ell + 1) \cdot \mathbf{C}_\ell^\mathrm{T} \cdot \mathrm{Cov}_{\ell}^{-1} \cdot \mathbf{C}_\ell
\end{equation}

Note that, in order to compute the S/N of the cross-correlation, we do not make use of the auto-correlation $C_\ell$'s appearing in the vector of equation \eqref{eq:vectors}. In this way we compute the S/N related only to the cross-correlation part, avoiding the contribution from auto-correlations which is likely to increase the S/N, if the auto-correlations are free of systematic effects. The rationale behind this treatment is to be conservative and assume that the cross-correlation signal is less prone to systematic effects compared to the auto-correlations, as it is indeed the case for other Large Scale Structure tracers.

More often, a cumulative Signal-to-Noise ratio $\Biggl(\dfrac{S}{N}\Biggl)(\ell < \ell_{\mathrm{max}})$ is considered. In both cases it is defined as
\begin{equation}
\biggl(\dfrac{S}{N}\biggl)(\ell < \ell_{\mathrm{max}}) = \sqrt{\sum_{\ell' = \ell_{\mathrm{min}}}^{\ell_{\mathrm{max}}} \biggl(\dfrac{S}{N}\biggl)^2(\ell ')} .
\label{eq:cum_SNR}
\end{equation}
In sections \ref{sec:Cl_SN_YESTOMO} and \ref{sec:Cl_SN_NOTOMO} we show the results for the two S/N calculations. Note that in the non tomographic case, characterized by one single redshift bin, the two estimates coincide.

\subsection{Tomographic case}
\label{sec:Cl_SN_YESTOMO}

We cross-correlate all the considered galaxy tracers (i.e. galaxies with $\psi > 10, 100, 300  M_\odot/$yr) with all the GWs tracers (i.e. BH-BH, BH-NS, NS-NS) along three or four redshift bins $z_{i,j}$. The number of bins and their ranges differ from case to case due to the different redshift ranges in which these tracers can be defined (see section \ref{section:tracers}). In table \ref{tab:z_tomo} we provide the redshift binning considered for each probe. In this section we provide results for the exemplificative case of $f_{\rm sky}=0.7$ and up to a maximum multipole of $\ell_{\mathrm{max}}=100$, corresponding to the best angular resolution reachable by ET (see e.g.,~\cite{klimenko:ET_lmax}). The angular power spectra for all the tracers combinations are shown in Appendix \ref{app:Cls}.  

\begin{table}
\centering
\begin{tabular}{cc|ccc|}%{ll|lll|}
\cline{3-5}
                        &                         &                                                                                                                                               & GW                                                                                                                                           &                                                                                                       \\ \cline{3-5} 
                        &                         & \multicolumn{1}{c|}{BH-BH}                                                                                                                    & \multicolumn{1}{c|}{BH-NS}                                                                                                                    & NS-NS                                                                                                 \\ \hline
\multicolumn{1}{|c|}{}  & $\psi > 10 M_\odot/$yr  & \multicolumn{1}{c|}{\begin{tabular}[c]{@{}c@{}}$0 \leq z_1 \leq 1$ \\ $1 \leq z_2 \leq 2$\\$ 2\leq z_3 \leq 3$\\$ 3\leq z_4 \leq 6$\end{tabular}}   & \multicolumn{1}{|c|}{\begin{tabular}[c]{@{}c@{}}$0 \leq z_1 \leq 1 $\\$ 1 \leq z_2 \leq 2$\\$ 2\leq z_3 \leq 3$\\$ 3\leq z_4 \leq 6$\end{tabular}}   & \begin{tabular}[c]{@{}c@{}}$0 \leq z_1 \leq 1 $\\$ 1 \leq z_2 \leq 2$\\$ 2\leq z_3 \leq 3$\end{tabular}   \\ \cline{2-5} 
\multicolumn{1}{|c|}{g} & $\psi > 100 M_\odot/$yr & \multicolumn{1}{c|}{\begin{tabular}[c]{@{}c@{}}$0.5 \leq z_1 \leq 1 $\\$ 1 \leq z_2 \leq 2$\\$ 2\leq z_3 \leq 3$\\$ 3\leq z_4 \leq 6$\end{tabular}} & \multicolumn{1}{c|}{\begin{tabular}[c]{@{}c@{}}$0.5 \leq z_1 \leq 1 $\\$ 1 \leq z_2 \leq 2$\\$ 2\leq z_3 \leq 3$\\$ 3\leq z_4 \leq 6$\end{tabular}} & \begin{tabular}[c]{@{}c@{}}$0.5 \leq z_1 \leq 1 $\\$ 1 \leq z_2 \leq 2$\\$ 2\leq z_3 \leq 3$\end{tabular} \\ \cline{2-5} 
\multicolumn{1}{|c|}{}  & $\psi > 300 M_\odot/$yr & \multicolumn{1}{c|}{\begin{tabular}[c]{@{}c@{}}$1 \leq z_1 \leq 2 $\\$ 2 \leq z_2 \leq 3$\\$ 3\leq z_3 \leq 6$\end{tabular}}                      & \multicolumn{1}{c|}{\begin{tabular}[c]{@{}c@{}}$1 \leq z_1 \leq 2 $\\$ 2 \leq z_2 \leq 3$\\$ 3\leq z_3 \leq 6$\end{tabular}}                      & \begin{tabular}[c]{@{}c@{}}$1 \leq z_1 \leq 2 $\\$ 2 \leq z_2 \leq 3$\end{tabular}                      \\ \hline
\end{tabular}
\caption{Redshift binning for the tomographic case.}
\label{tab:z_tomo}
\end{table}

\begin{figure}[H]
    \centerline{
		\subfloat{
			\includegraphics[width=1.0\linewidth]{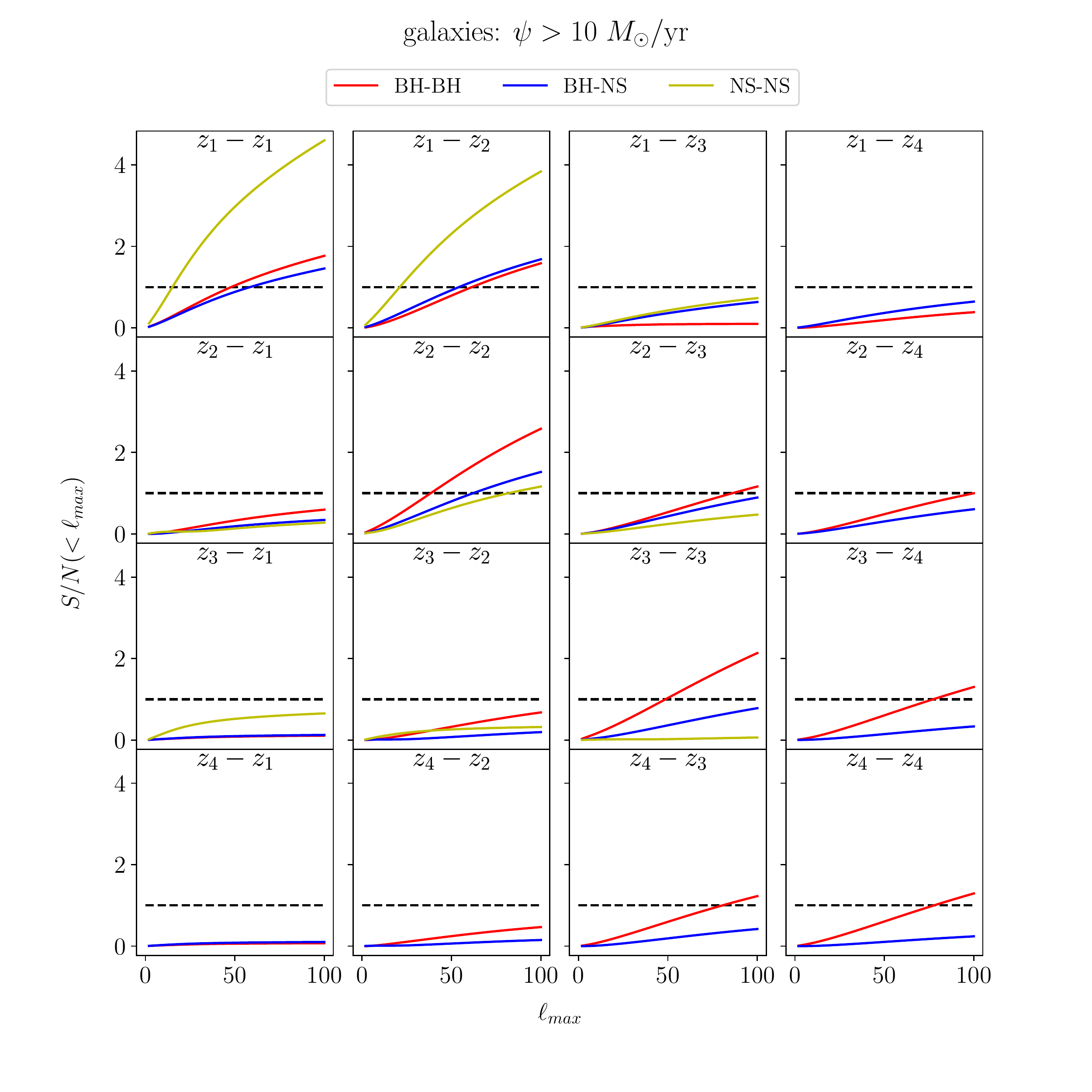}}}
	\caption{Cumulative Signal-to-Noises ${\rm S/N}(<\ell_{\mathrm{max}})$ (equations \eqref{eq:SNR_1} and \eqref{eq:cum_SNR}) for the cross-correlations cases between galaxies with $\psi > 10 M_\odot/$yr with all three types of GWs signals (BH-BH in red, BH-NS in blue, NS-NS in yellow). Horizontal dashed lines correspond to ${\rm S/N}(<\ell_{\mathrm{max}}) = 1$. The plot refers to $T_{\mathrm{obs}}=1$yr and $f_{\mathrm{sky}}=0.7$.}
	\label{fig:SN_cumulative_g0_rho_abv_8_YESTOMO}
\end{figure}

\begin{figure}[H]
    \centerline{			
		\subfloat{
			\includegraphics[width=1.0\linewidth]{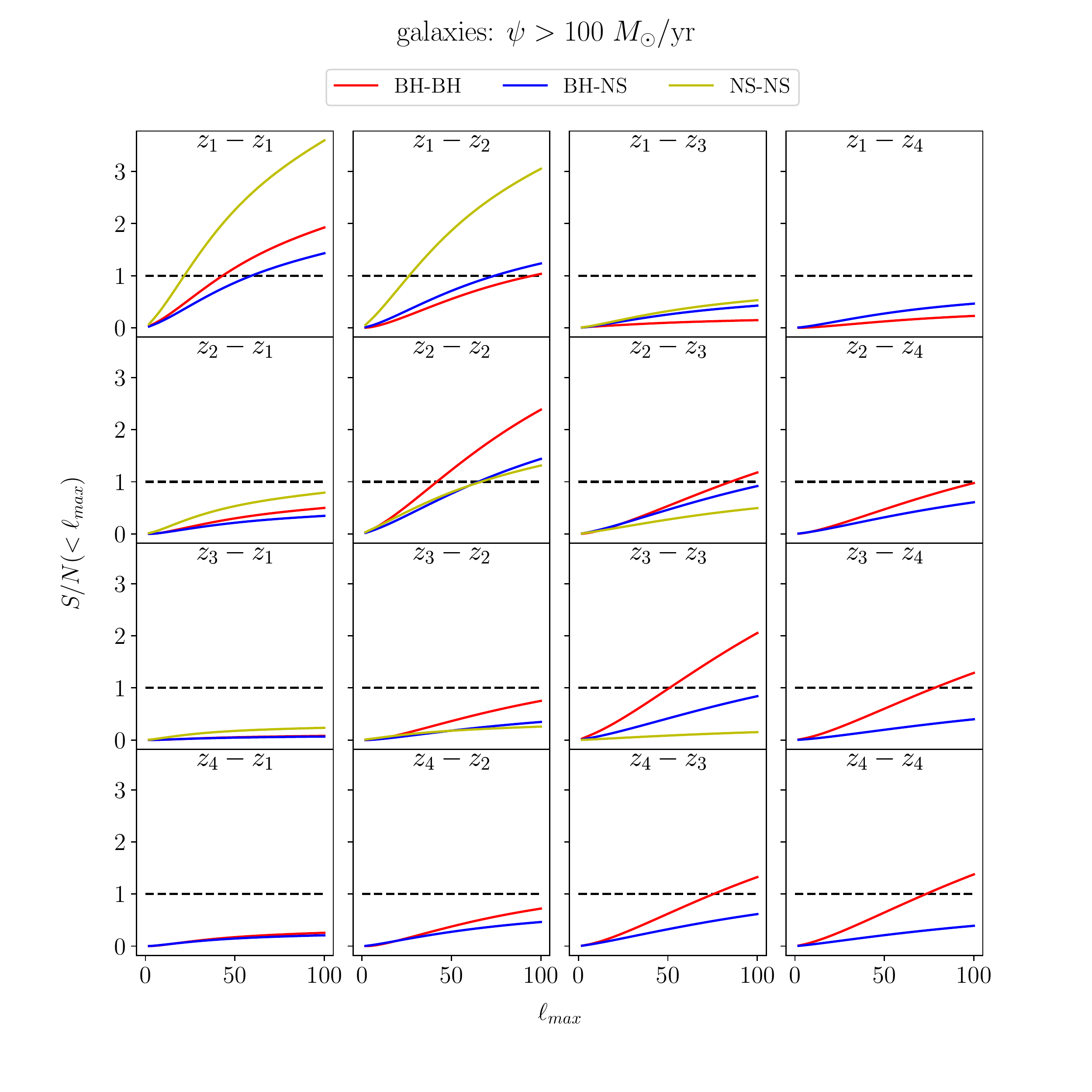}}}
	\caption{Cumulative Signal-to-Noises ${\rm S/N}(<\ell_{\mathrm{max}})$ (equations \eqref{eq:SNR_1} and \eqref{eq:cum_SNR}) for the cross-correlations cases between galaxies with $\psi > 100 M_\odot/$yr with all three types of GWs signals (BH-BH in red, BH-NS in blue, NS-NS in yellow). Horizontal dashed lines correspond to ${\rm S/N}(<\ell_{\mathrm{max}}) = 1$. The plot refers to $T_{\mathrm{obs}}=1$yr and $f_{\mathrm{sky}}=0.7$.}
	\label{fig:SN_cumulative_g1_rho_abv_8_YESTOMO}
\end{figure}

\begin{figure}[H]
    \centerline{
		\subfloat{
			\includegraphics[width=1.0\linewidth]{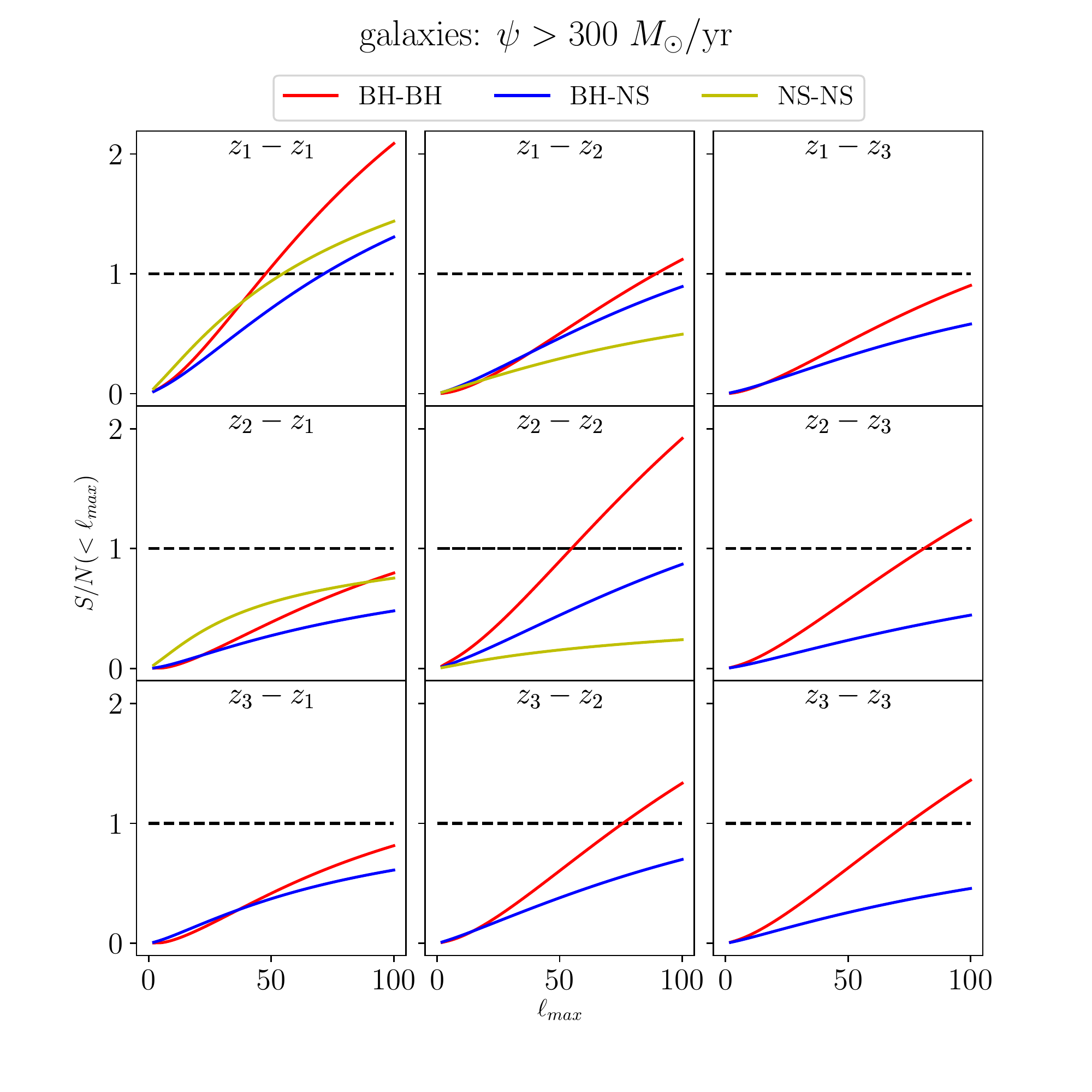}}}
	\caption{Cumulative Signal-to-Noises ${\rm S/N}(<\ell_{\mathrm{max}})$ (equations \eqref{eq:SNR_1} and \eqref{eq:cum_SNR}) for the cross-correlations cases between galaxies with $\psi > 300 M_\odot/$yr with all three types of GWs signals (BH-BH in red, BH-NS in blue, NS-NS in yellow). Horizontal dashed lines correspond to ${\rm S/N}(<\ell_{\mathrm{max}}) = 1$. The plot refers to $T_{\mathrm{obs}}=1$yr and $f_{\mathrm{sky}}=0.7$.}
	\label{fig:SN_cumulative_g2_rho_abv_8_YESTOMO}
\end{figure}

In figures \ref{fig:SN_cumulative_g0_rho_abv_8_YESTOMO}, \ref{fig:SN_cumulative_g1_rho_abv_8_YESTOMO}, \ref{fig:SN_cumulative_g2_rho_abv_8_YESTOMO} we provide the cumulative S/N computed from the first method (i.e. applying equation \eqref{eq:cum_SNR} to \eqref{eq:SNR_1}) for all combinations of tracers and redshift bins. Each subplot shows the estimates for one specific galactic tracer, with every $\mathrm{GW}$ tracer. The following notation regarding the redshift binning is adopted: given a g$\times$GW couple, the notation $z_i - z_j$ means that we are cross-correlating galaxies in bin $z_i$ with GWs in bin $z_j$. Note that each subplot refers to cross-correlations between redshift bins $z_i - z_j$, but each $z_{i,j}$ can actually be different between different tracers according to table \ref{tab:z_tomo}. For this reason, curves with different colors should not be directly compared to one another, since they refer to different ranges.

As for the S/N values, it can be seen that in several cases ${\rm S/N}(<\ell_{\mathrm{max}}) > 1$. In particular: \textit{i)} for all galaxy tracers, the highest ${\rm S/N}(<\ell_{\mathrm{max}})$ are found for correlations among the same redshift bins; \textit{ii)} cross-correlations between distant redshift bins are those providing a lower ${\rm S/N}(<\ell_{\mathrm{max}})$: even though effects such as lensing can induce even a strong correlation between distant objects, in many of the cases considered here it is not enough to strengthen the ${\rm S/N}$; \textit{iii)} considering correlations among the same bins, the ${\rm S/N}(<\ell_{\mathrm{max}})$ are strongly sensitive to the amount of detected sources: for large redshift values (bins $z_{2,3,4}$) they are always higher in the BH-BH case, followed by BH-NS and eventually by NS-NS. Indeed, the BH-BH case corresponds to a higher number of merging events (as can be seen by looking at its redshift distribution of the left panel of figure \ref{fig:specifics_gw}). This provides a smaller amount of shot noise, which contributes to making the ${\rm S/N}$ of equation \eqref{eq:SNR_1} higher. On the other hand, at low redshift (bin $z_1$) the redshift distribution of NS-NS mergers is significantly higher than the others, which is reflected in a ${\rm S/N}(<\ell_{\mathrm{max}})$ which is often bigger; \textit{iv)} in analogy with the previous point, at fixed GW tracer the higher is the cut in SFR, the lower is the S/N, reflecting the smaller number of galaxies considered.

\begin{figure}[H]
    \centerline{
		\subfloat{
			\includegraphics[width=1.0\linewidth]{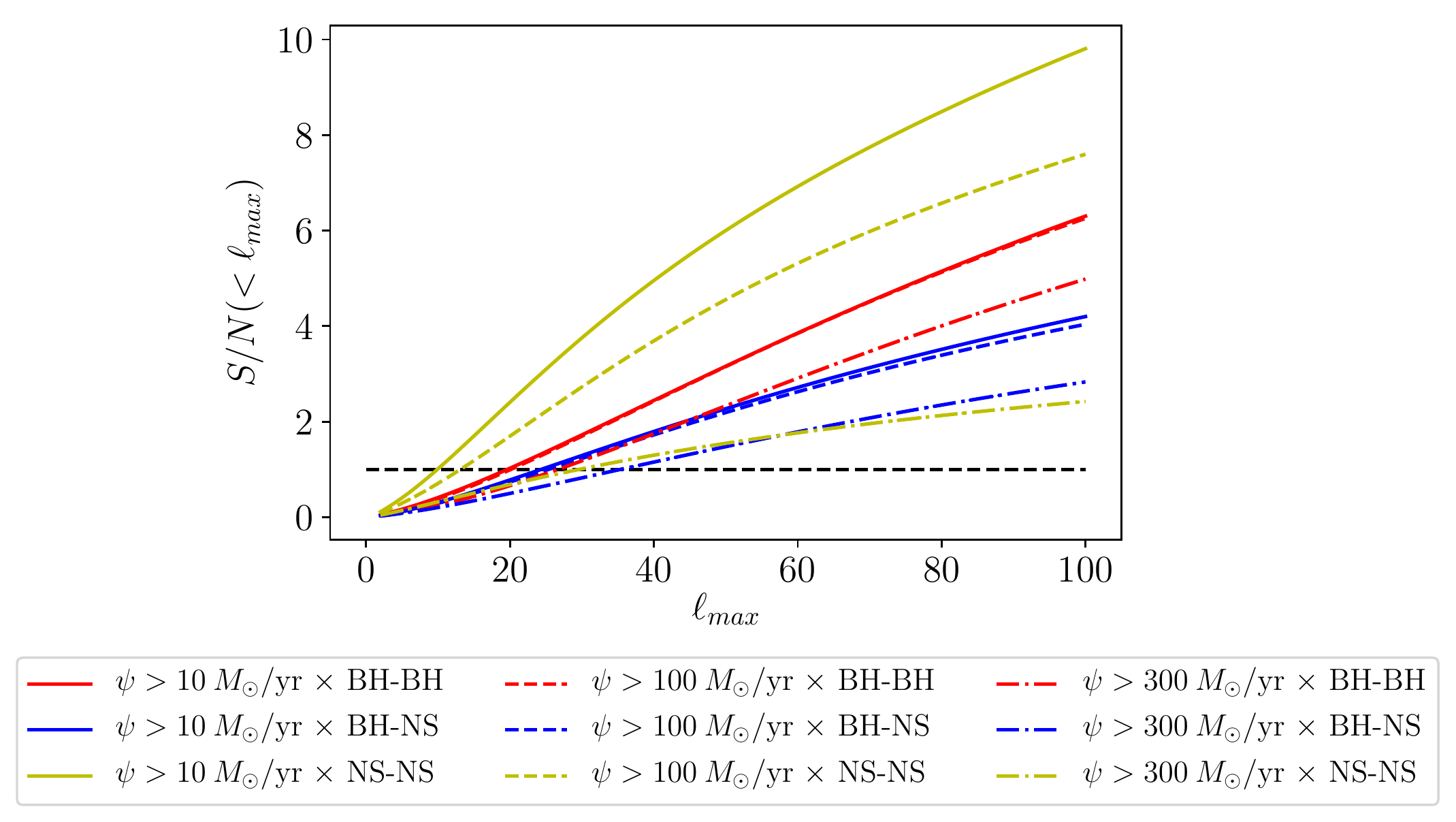}}}
	\caption{Cumulative Signal-to-Noises ${\rm S/N}(<\ell_{\mathrm{max}})$ for all cross-correlation cases (equations \eqref{eq:SNR_2} and \eqref{eq:cum_SNR}). The horizontal dashed line corresponds to ${\rm S/N}(<\ell_{\mathrm{max}}) = 1$. Line-styles refer to galaxies ($\psi > 10 M_\odot/$yr in full line, $\psi > 100 M_\odot/$yr in dashed line, $\psi > 300 M_\odot/$yr in dotted-dashed line) while colors refer to gravitational waves (BH-BH in red, BH-NS in blue, NS-NS in yellow). The plot refers to $T_{\mathrm{obs}}=1$yr and $f_{\mathrm{sky}}=0.7$.}
	\label{fig:SN_TOT_CUM_TOMO}
\end{figure}

The cumulative S/N for the whole probe, computed applying equation \eqref{eq:cum_SNR} to \eqref{eq:SNR_2}, is shown in figure \ref{fig:SN_TOT_CUM_TOMO}. The line-styles refer to a specific galaxy tracer, while the colors distinguish between GW types (as indicated in the legend). It can be seen that generally it reaches values above unity for all the tracers combinations. Note that the cross-correlation signal overcomes the noise already for relatively low multipoles, at around $\ell_{\mathrm{max}}\sim10-40$. The S/N is particularly high especially for the $\psi>10 M_\odot/\rm yr\times$ NS-NS case, where it reaches a value of $\sim 10$. The NS-NS case is also the most dependent on the chosen SFR cut, because the peak of the detected NS-NS distribution is at rather low redshift ($z\leq 1$), where the distribution of highly star forming galaxies tends to fall down (see figures \ref{fig:specifics_g} and \ref{fig:specifics_gw}, left panels). We stress again that the different cases should not be directly compared since they refer to different redshift ranges, depending on the tracers characteristic intervals. All in all, cross-correlations between the treated tracers, adopting a tomographic approach, can be informative given the rather high S/N ratio values.

\subsection{Non-tomographic case}
\label{sec:Cl_SN_NOTOMO}

In this subsection we compute the cross-correlations between our tracers without a tomographic approach. This is done to see how the measured cross-correlation signal can differ when squeezing all the detected sources into one single bin, without a sliced tomographic approach. As for the previous case, the redshift ranges of all the tracers combinations differ for each combination. In table \ref{tab:z_no_tomo} we provide the redshift ranges considered for each probe.

\begin{table}
\centering
\begin{tabular}{cc|ccc|}
\cline{3-5}
                        &                         &                                           & GW                                       &                      \\ \cline{3-5} 
                        &                         & \multicolumn{1}{c|}{BH-BH}                & \multicolumn{1}{c|}{BH-NS}                & NS-NS                \\ \hline
\multicolumn{1}{|c|}{}  & $\psi > 10 M_\odot/$yr  & \multicolumn{1}{c|}{$ 0 \leq z \leq 6$}   & \multicolumn{1}{c|}{$ 0 \leq z \leq 6$}   & $ 0 \leq z \leq 3$   \\ \cline{2-5} 
\multicolumn{1}{|c|}{g} & $\psi > 100 M_\odot/$yr & \multicolumn{1}{c|}{$ 0.5 \leq z \leq 6$} & \multicolumn{1}{c|}{$ 0.5 \leq z \leq 6$} & $ 0.5 \leq z \leq 3$ \\ \cline{2-5} 
\multicolumn{1}{|c|}{}  & $\psi > 300 M_\odot/$yr & \multicolumn{1}{c|}{$ 1 \leq z \leq 6$}   & \multicolumn{1}{c|}{$ 1 \leq z \leq 6$}   & $ 1 \leq z \leq 3$   \\ \hline
\end{tabular}
\caption{Redshift binning for the non-tomographic case.}
\label{tab:z_no_tomo}
\end{table}

\begin{figure}[H]
    \centerline{
		\subfloat{
			\includegraphics[width=1.05\linewidth]{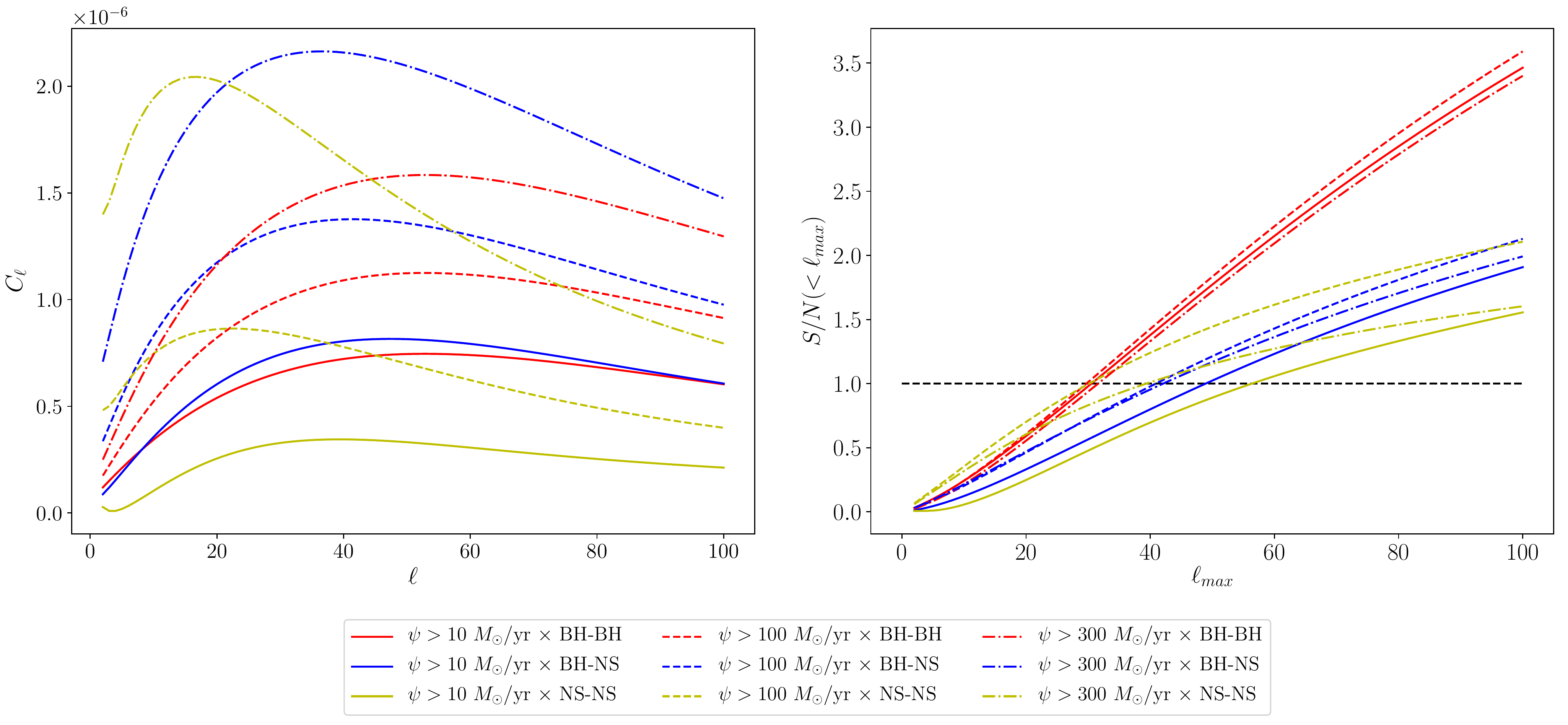}}}
	\caption{\textit{Left:} angular power spectra $C_\ell$'s for all cross-correlations cases. \textit{Right:} cumulative Signal-to-Noises ${\rm S/N}(<\ell_{\mathrm{max}})$ for all cross-correlations cases. The horizontal dashed line corresponds to ${\rm S/N}(<\ell_{\mathrm{max}}) = 1$. Line-styles refer to galaxies ($\psi > 10 M_\odot/$yr in full line, $\psi > 100 M_\odot/$yr in dashed line, $\psi > 300 M_\odot/$yr in dotted-dashed line) while colors refer to gravitational waves (BH-BH in red, BH-NS in blue, NS-NS in yellow). The plot refers to $T_{\mathrm{obs}}=1$yr and $f_{\mathrm{sky}}=0.7$.}
	\label{fig:Cls_SN_NOTOMO}
\end{figure}

In figure \ref{fig:Cls_SN_NOTOMO} (left panel) we show the angular power spectra for all combinations of tracers. As in figure \ref{fig:SN_TOT_CUM_TOMO}, line-styles/colors refer to galaxies/GWs. It can be seen that the power spectra are higher when galaxies with a higher SFR are considered, as logically expected, since a higher SFR leads to a larger absolute number of remnants, and so to a larger amount of merging pairs.

In figure \ref{fig:Cls_SN_NOTOMO} (right panel) we provide the cumulative ${\rm S/N}(<\ell_{\mathrm{max}})$ for each probe. We stress again that the S/N, computed applying equation \eqref{eq:cum_SNR} to equations \eqref{eq:SNR_1} and \eqref{eq:SNR_2}, are in this case coincident. It can be seen that for an $\ell_{\mathrm{max}}$ large enough a ${\rm S/N}(<\ell_{\mathrm{max}})>1$ is always reached for any tracer combination. The cross-correlation signal overcomes the noise at around $\ell_{\mathrm{max}}\sim20-40$. The NS-NS contribution in this case is lower with respect to the tomographic approach because, even if their shot noise is small, their $C_\ell$'s values are also small, due to the fact that NS-NS mergers are mostly seen at low redshifts making their distribution rather different with respect to the galaxies one.

By comparing figure \ref{fig:SN_TOT_CUM_TOMO} and figure \ref{fig:Cls_SN_NOTOMO} (right panel) it is possible to gauge the fact that assuming a tomographic approach is indeed an advantage, since the extra (radial) information provided contributes to build a stronger S/N.

\section{Cross-correlations as astrophysical probe}\label{sec:discerning_scenarios}

In this section we are going to discuss about the possibility to exploit $\mathrm{GW \times LSS}$ cross-correlations to compare and test the validity of different astrophysical scenarios concerning the formation, evolution and merging of COs binaries. Given the uncertainties in the astrophysics and the enormous modeling possibilities, it is at the moment unlikely to be able to unequivocally determine the validity of one specific combination of prescriptions with respect to any possible other. For this reason, our approach does not aim at excluding or validating specific models, whereas it looks at the possibility to apply this methodology to some general proof-of-concept cases, leaving the chance to use this formalism to whom is interested in applying it for specific tests. In fact, it is worth stressing again that the application we presented in this section is not the only possible one. Any other astrophysical formulation and modeling that influences the estimate of redshift distributions and/or biases of the considered tracers can in principle be addressed. This can cover a wide range of possibilities, from the stellar modeling (especially regarding the estimation of the CO mass and the constrain of all those processes influencing it) to galaxy evolution and SFRs calculation methods.

To compare two models (which we describe in section \ref{sec:Z_dependency}) and investigate the possibility to discern them, we make use of a $\Delta \chi^2$ statistics, whose formalism and results are presented in section \ref{sec:chi2_metallicity}.

\subsection{Compared astrophysical models}\label{sec:Z_dependency}
The two cases that we compare in this study differ in the treatment of the effect of the galaxy metallicity on binary evolution. The metallicity value of the ZAMS stars constituting the binary, as already mentioned in section \ref{section:merging_rates}, is fundamental in order to determine the subsequent binary evolution. The value of metallicty has a strong impact both on the COs mass and on the number of COs merging per unit of star forming mass. In particular the higher is the metallicity the lower is the chirp mass and the number of merging binaries per unit of star forming mass. The physical reasons of these dependencies are well explained in e.g.,~\cite{belczynski:massivebhsmergers},~\cite{dominik+12},~\cite{chruslinska+18}.

In section \ref{section:merging_rates} and throughout this paper we accounted for the differences in the compact remnant masses due to the different metallicities present in galaxies through the factor $  dp/d\log_{10}\mathcal{M}_{\bullet\bullet}\,(\mathcal{M}_{\bullet\bullet}|Z)$ in equation \eqref{eq|rmerge}. However, as explained in section \ref{section:merging_rates}, we did not consider the possible dependence on metallicity of the merging efficiency due to binary evolution effects. This is translated in the fact that we have chosen a factor $  f_{\rm  eff}$ independent on metallicity, whose value was determined by normalizing the merging rates to the local value constrained by the LIGO/Virgo team.

In this specific section we change approach, aiming to address the issue of the possibility to distinguish, through $\mathrm{GW\times LSS}$ cross-correlations, between different astrophysical models of stellar and binary evolution. Therefore, we compare our precedent results obtained with a constant $  f_{\rm  eff}$ (we refer to it as "benchmark case") with one of the models with a metallicity dependent $  f_{\rm  eff}\,(Z)$. As an example we choose the reference model presented in~\cite{chruslinska+18}. The metallicity dependence of the number of merging events per unit star forming mass of such a model is shown in figure 1 of~\cite{chruslinska+19a}, thin lines (we refer to it as "Z-dependent case"). However, the reader should keep in mind that we are considering only the shape of this factor as a function of metallicity, the normalization is still fixed by the local merging rate given by the LIGO/Virgo teams. The equation to compute the merging rates now becomes:
\begin{equation}
\begin{split}
    \frac{  d^2\dot{N}_{  merge}}{  dz\,d\mathcal{M}_{\bullet\bullet}}(t,\mathcal{M}_{\bullet\bullet})=&  \tilde{a}\frac{  dV}{  dz\,(1+z)}\int  dt_{  d}\frac{  dp}{  dt_{  d}}\int\log_{10}\psi\frac{  d^2N(\log_{10}\psi|t-t_{  d})}{  d\log_{10}\psi\,dV}\psi\times\\
    &\times\int  d\log_{10} Z\frac{  dp}{  d\log_{10} Z}(\log_{10} Z|t-t_{  d},\psi)\,  f_{\rm  eff}(Z)\,\frac{  dp}{  d\mathcal{M}_{\bullet\bullet}}(\mathcal{M}_{\bullet\bullet}|Z)\, ,
\end{split}
\label{eq|rmergemetal}
\end{equation}
which is identical to equation \eqref{eq|rmerge}, except for the metallicity dependent $f_{\rm  eff}$ factor, which now takes part in the integration. The factor $  \tilde{a}$ instead guarantees the normalization to the LIGO/Virgo local values mentioned above.

Thus, we are comparing a simple case (the benchmark one), where the dependence on metallicity enters only in the remnant mass distribution, but not in the number of merging binaries per unit of star forming mass, with a more realistic model (the Z-dependent case) where this latter dependence is included. The differences in the shape of the merging rates between the two cases enters in the computation of the $C_\ell$'s allowing the possibility to distinguish between them. On the other hand differences in the absolute number of sources affects the shot noise of the two cases, in particular the number of merging events of the benchmark case tends to be, on average, higher with respect to the Z-dependent case, resulting in a lower shot noise.

Clearly this is a case study to check whether it is possible to detect this metallicity imprinting through $\mathrm{GW \times LSS}$ cross correlations, but this technique could be in principle pursued in more refined studies to test different astrophysical models. One of the reasons of exploiting the cross-correlation formalism to test these two cases is given by the fact that, since SFR and metallicity are interconnected parameters, a dependence on the metallicity of the efficiency with which COs binaries merge will (non-trivially) be expressed also as a dependence on the SFR, causing GWs mergers to correlate differently with galaxies of different SFR values. We finally remark that we are going to compare these two models only for the BH-BH merging case for two main reasons: the first one is that the metallicity dependence of the number of merging events is stronger for the BH-BH case (see~\cite{chruslinska+19a}), the second one is that BH-BH events are much more frequent with respect to the other types of merging: this, as already seen, reduces the shot noise enhancing the S/N of the cross-correlation.

\subsection{Theoretical forecasts}
\label{sec:chi2_metallicity}

In this section we provide the forecasts for discerning the benchmark scenario from the metallicity dependent one. We make use of a $\Delta \chi^2$ statistics to evaluate a S/N, whose value (above/below unity) can provide information on how different the two models (one called as Fiducial and the other as Alternative) are. Following the same approach of~\cite{Scelfo_2018} we define a S/N as: 
\begin{equation}
\left(\frac{S}{N}\right)^2 \sim \Delta\chi^2 := f_{\mathrm{sky}}\sum_2^{\ell_\mathrm{max}} (2\ell+1) (\mathbf{C}^\mathrm{Alternative}_\ell-\mathbf{C}^\mathrm{Fiducial}_\ell)^T \mathrm{Cov}^{-1}_\ell (\mathbf{C}^\mathrm{Alternative}_\ell-\mathbf{C}^\mathrm{Fiducial}_\ell),
\label{eq:delta_chi2}
\end{equation}
where $\mathbf{C}_\ell^{\mathrm{Fiducial/Alternative}}$ is a vector containing the $C_\ell$'S from the Fiducial/Alternative model, organized with the same logic of equation \eqref{eq:vectors} and where the $\mathrm{Cov}_\ell$ is the covariance matrix as in equation \eqref{eq:covariance_matrix_chi2}, built with the $C_\ell$'s of the fiducial model. Since the entries of the covariance matrix depend on which model is assumed as fiducial, the final forecasts also depend on this choice. For this reason, we computed S/N in both cases and compared them.

\begin{figure}[H]
    \centerline{
		\subfloat{
			\includegraphics[width=0.75\linewidth]{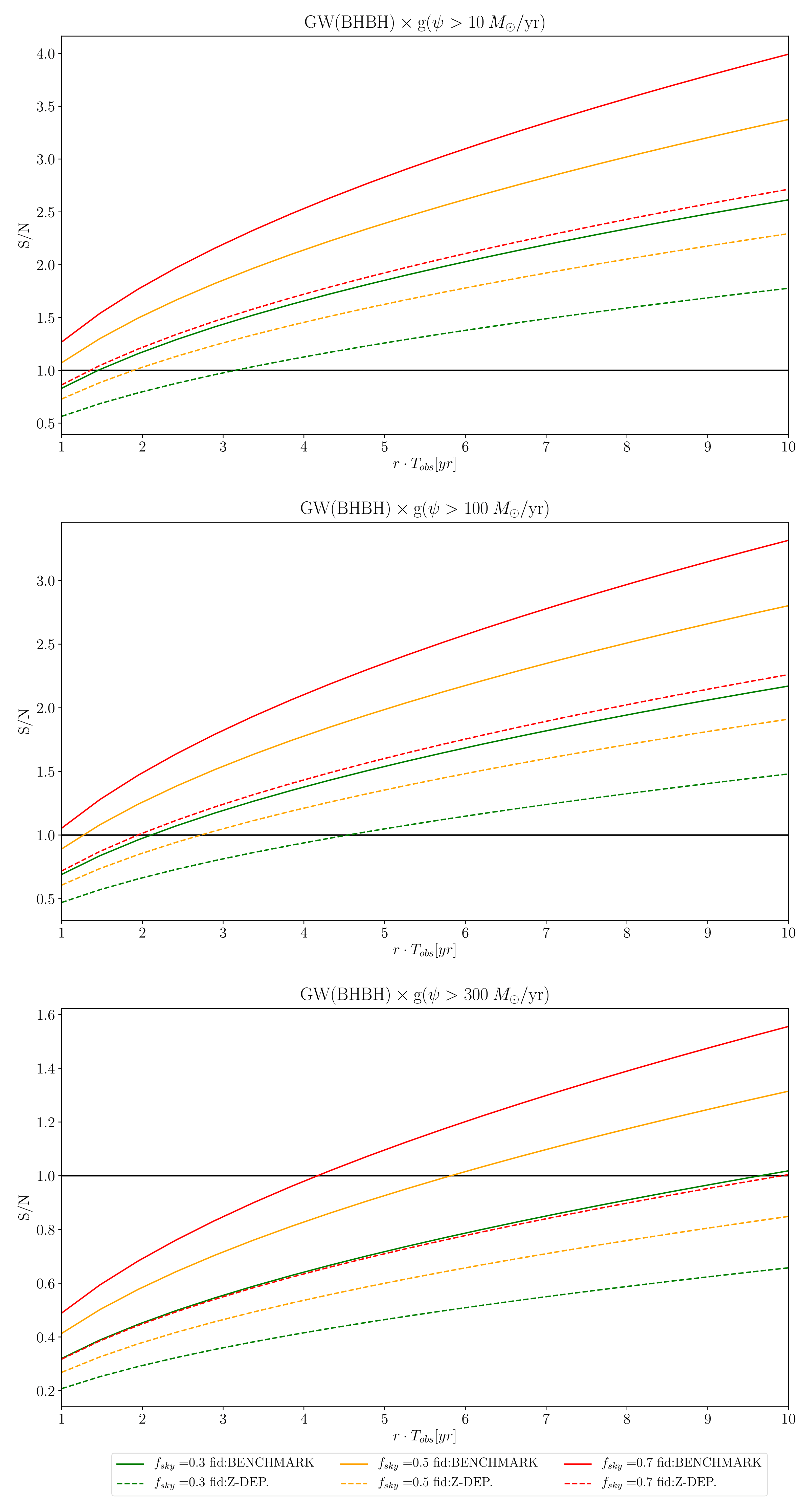}}}
	\caption{S/N from $\Delta \chi^2$ analysis (equation \eqref{eq:delta_chi2}) for discerning the two considered astrophysical scenarios. Galaxies with $\psi>10,100,300 M_{\odot}/$yr are considered (top, central and bottom panel respectively). Continuous/dashed lines refer to the benchmark/Z-dependent model assumed as fiducial. Colors refer to different values of $f_{\mathrm{sky}}$ as shown in legend.}
	\label{fig:chi2_all}
\end{figure}

In figure \ref{fig:chi2_all} we provide the S/N obtained considering galaxies with $\psi > 10, 100, 300 M_{\odot} /$yr. We show results for different observed sky-fractions $f_{\mathrm{sky}}$, observation times $T_{\mathrm{obs}}$ and for both models assumed as fiducial. More precisely, our results are expressed not only as a function of  $T_{\mathrm{obs}}$ (on the horizontal axis) but, instead, of the product $r \cdot T_{\mathrm{obs}}$. The quantity $r$ is a multiplicative fudge factor to the merging rate of GWs introduced to take into account any possible uncertainty in the modeling of this quantity. Note that $r$ and $T_{\mathrm{obs}}$ are degenerate: for example, observing for $T_{\mathrm{obs}}= 1\: \rm{yr}$ with a factor $r=2$ yields the same result as observing for $T_{\mathrm{obs}}= 2 \: \rm{yr}$ with a factor $r=1$. The case of $r=1$ corresponds to the scenario in which the models used here are the "true" ones. It is worth noticing that the $r$ factor has the same effect of the $\tilde{a}$ factor that quantifies the normalization to the local observed rate introduced in equation \ref{eq|rmergemetal}, since they both are multiplicative factors to the merger rate. For this reason, the $r$ factor can also be seen as absorbing the uncertainties on the local merging rates estimates.

First of all, it can be seen that when the benchmark model is assumed as fiducial, the forecasts are significantly better compared to the opposite (for fixed $f_{\mathrm{sky}}$ and $T_{\mathrm{obs}}$): this is due to the fact that this model predicts a higher number of GWs mergers, providing a smaller shot noise contribution. For analogous reasons, when comparing the panels in figure \ref{fig:chi2_all}, it can be seen that results are more optimistic when considering galaxies with $\psi>10 M_{\odot} /$yr: the higher the number of sources (galaxies in this case) the better the results. This is also reflected on the fact that the case with galaxies of $\psi>300 M_{\odot} /$yr is the most pessimistic.

Looking in detail at each of the figures, we can see that in the $\psi>10 M_{\odot} /$yr case a S/N above unity can be reached in a relatively short time: even for small observed fractions of the sky (e.g., $f_{\mathrm{sky}}=0.3$) not more than 3 years of observation would be required to marginally distinguish the two scenarios. Looking instead at the most pessimistic case, in which only galaxies with $\psi>300 M_{\odot} /$yr are considered, approximately 5 to 10 years of observation would be required to reach ${\rm S/N} \sim 1$ in the case of benchmark model assumed as fiducial. A much higher observation time (at least above 10 years) is required when assuming the $Z$-dependent scenario as fiducial. The case of $\psi>100 M_{\odot} /$yr lies in between, with a still fairly optimistic prediction.

All in all our results are rather promising, especially when considering galaxies with $\psi>10\,\rm M_\odot/yr$ and $\psi>100\,\rm M_\odot/yr$: if the benchmark case is the fiducial one, deviations from it can be detected after just $2\rm \: yr$ of observational time. If the $Z$-dependent case is the fiducial, we should be able to detect variations from it in $\lesssim 5\rm \: yr$ of observations. For highly star forming galaxies with $\psi>300\,\rm M_\odot/yr$ instead, some more time is required to distinguish the two models. Still an observational time $\lesssim 10\rm \: yr$ is enough if the benchmark case is considered as fiducial.

Finally, we stress again that this forecast does not aim at testing or excluding any of the two models considered here. It aims instead at showing how different astrophysical prescriptions (such as a $Z$ dependence on the $  f_{\mathrm{eff}}$ factor) could in principle be distinguished through the cross-correlation formalism, contributing in tackling different astrophysical issues.

\section{Conclusions}\label{sec:conclusions}

In this work we have expanded the investigation in the field of the cross-correlations between resolved GWs signals and LSS tracers. We worked in the harmonic space with the number counts angular power spectra. The two categories of tracers we considered consist in resolved GWs events from BH-BH, BH-NS, NS-NS mergers detectable by the Einstein Telescope and actively star-forming galaxies with SFR cuts of $\psi > 10, \: 100, \: 300 \: M_\odot/$yr. We characterized them with their redshift distributions, bias and magnification bias values, presenting a detailed description of the computation of these quantities. We stress again that both the SFRF and the GWs distributions derive from the same type of sources (galaxies) but trace them in a different way, since GW signals depend not only on the galaxy SFR, but also on the galaxy metallicity and on stellar and binary evolution prescriptions. For this reasons we kept into account all the aforementioned elements in this work. Cross-correlating the same sources via two different messengers can help not only in alleviating systematics but also in enhancing the amount of astrophysical information encoded in the signal.

In our analysis we took into account all lensing and all general relativistic contributions in the computation of the observed number counts fluctuations and we extended the basis for future works regarding the $\mathrm{GW \times LSS}$ cross-correlations with a more robust theoretical astrophysical background.

After computing the number counts angular power spectra for all the combinations of our $\mathrm{GW \times LSS}$ tracers, we estimated Signal-to-Noise ratios in order to forecast the detectability of the cross-correlation signal. We have  considered both tomographic and non-tomographic approaches. Our results show that in several scenarios it is possible to reach a Signal-to-Noise ratio higher than unity, whereas it is not always the case for cross-correlations between distant redshift bins (in the tomographic case) or with a low number of observed objects. In addition, the total cumulative Signal-to-Noise ratios for each of the probes considered in this work are in turn quite optimistic. Even though each of the many considerable surveys (such as ALMA~\cite{Wootten:ALMA}, JWST~\cite{gardner:JWST}, EMU~\cite{Norris:EMU}, SKA~\cite{Maartens:SKA} and many others) will have its own specifics, this work still provides a general possibility to gauge the cross-correlations efficacy.

Finally, we have investigated the possibility of exploiting the $\mathrm{GW \times LSS}$ formalism to compare and test possible scenarios in the astrophysical modeling of GWs events. In particular we considered a proof-of-concept case in which we made use of a $\Delta \chi^2$ statistics to compare the cross-correlation signal obtained by modeling the COs merging efficiency with a specific metallicity dependency with respect to a benchmark signal obtained neglecting this dependency. We have showed that in principle, given enough individual objects observed (i.e. enough observation time, observed fraction of the sky, etc.) a metallicity dependency feature could be discerned from the flat benchmark case. This is another step in the promising multi-tracers field and towards its astrophysical applications for future works to come.

\acknowledgments
This work has been partially supported by PRIN MIUR 2017 prot. 20173ML3WW 002, `Opening the ALMA window on the cosmic evolution of gas, stars and supermassive black holes'. A.L. acknowledges the MIUR grant `Finanziamento annuale individuale attivit\'a base di ricerca' and the EU H2020-MSCA-ITN-2019 Project 860744
`BiD4BEST: Big Data applications for Black hole Evolution STudies'.
M.V. and G.S. are supported by INDARK PD51 INFN grant. M.V. is
also supported by ASI-INAF grant n.2017-14-H.0. We are thankful to Nicola Bellomo, Jos\'e Luis Bernal and Alvise Raccanelli for critical reading and helpful suggestions on an earlier version of this manuscript. L.B. acknowledges Martyna Chruslinska for helpful discussions. We are thankful to the anonymous referee for thoughtful evaluation and helpful suggestions given to improve our manuscript.

\appendix
\section{Relativistic number counts}
\label{app:deltas}
We provide here the full expression for the relativistic number counts effects written in equation \eqref{eq:numbercount_fluctuation}:
\begin{equation}
\begin{aligned}
\Delta_\ell^\mathrm{den}(k,z) &= b_X \delta(k,\tau_z) j_\ell,	\\
\Delta_\ell^\mathrm{vel}(k,z) &=  \frac{k}{\mathcal{H}}j''_\ell  V(k,\tau_z) + \left[(f^\mathrm{evo}_X-3)\frac{\mathcal{H}}{k}j_\ell + \left(\frac{\mathcal{H}'}{\mathcal{H}^2}+\frac{2-5s_X}{r(z)\mathcal{H}}+5s_X-f^\mathrm{evo}_X\right)j'_\ell \right]  V(k,\tau_z),	\\
\Delta_\ell^\mathrm{len}(k,z) &= \ell(\ell+1) \frac{2-5s_X}{2} \int_0^{r(z)} dr \frac{r(z)-r}{r(z) r} \left[\Phi(k,\tau_z)+\Psi(k,\tau_z)\right] j_\ell(kr),	\\
\Delta_\ell^\mathrm{gr}(k,z)  &= \left[\left(\frac{\mathcal{H}'}{\mathcal{H}^2}+\frac{2-5s_X}{r(z)\mathcal{H}}+5s_X-f^\mathrm{evo}_X+1\right)\Psi(k,\tau_z) + \left(-2+5s_X\right) \Phi(k,\tau_z) + \mathcal{H}^{-1}\Phi'(k,\tau_z)\right] j_\ell + \\
&+ \int_0^{r(z)} dr \frac{2-5s_X}{r(z)} \left[\Phi(k,\tau)+\Psi(k,\tau)\right]j_\ell(kr) , \\
&+ \int_0^{r(z)} dr \left(\frac{\mathcal{H}'}{\mathcal{H}^2}+\frac{2-5s_X}{r(z)\mathcal{H}}+5s_X-f^\mathrm{evo}_X\right)_{r(z)} \left[\Phi'(k,\tau)+\Psi'(k,\tau)\right] j_\ell(kr).
\end{aligned}
\end{equation}
The meaning of the physical quantities written above is the following: $b_X$ is the bias parameter, $s_X$ is the magnification bias parameter, $f^\mathrm{evo}_X$ is the evolution bias parameter, $r$ is the conformal distance on the light cone, $\tau=\tau_0-r$ is the conformal time, $\tau_z=\tau_0-r(z)$, $j_\ell$, $j'_\ell=\frac{dj_\ell}{dy}$, $j''_\ell=\frac{d^2j_\ell}{dy^2}$ are the Bessel functions and their derivatives (evaluated at $y=kr(z)$ if not explicitly stated), $\mathcal{H}$ is the conformal Hubble parameter, the prime symbol $'$ stands for derivatives with respect to conformal time, $\delta$ is the density contrast in the comoving gauge, $V$ is the peculiar velocity, $\Phi$ and $\Psi$ are Bardeen potentials.

\section{Angular power spectra for the tomographic case}\label{app:Cls}
We provide here the angular power spectra $C_\ell$'s for all combinations of tracers in the tomographic approach described in section \ref{sec:Cl_SN_YESTOMO}. Note that each subplot reports cross-correlations of redshift bins $z_i - z_j$, but each $z_{i,j}$ can actually be different between different tracers according to table \ref{tab:z_tomo}. Some cases present power spectra with downward spikes: this is because we plotted the absolute values of the $C_\ell$'s and the spikes simply correspond to change-of-sign multipoles (the reader interested in the non-surprising possibility of having negative $C_\ell$'s can read e.g., ref.~\cite{DiDio:negative_Cls}).
\begin{figure}[H]
    \centerline{
		\subfloat{
			\includegraphics[width=1.0\linewidth]{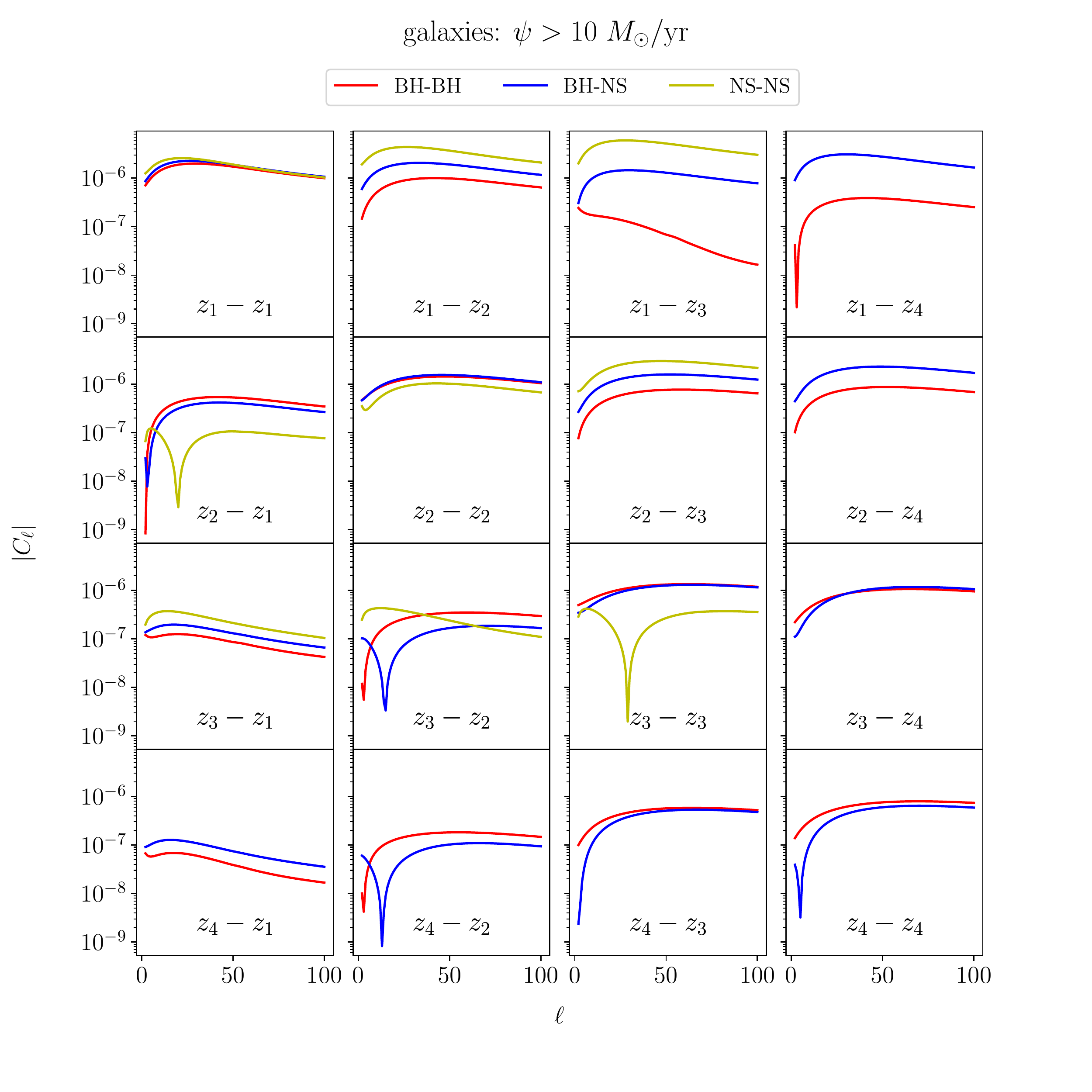}}}
	\caption{Angular power spectra $C_\ell$'s (absolute values) for the cross-correlations tomographic cases between galaxies with $\psi > 10 M_\odot/$yr with all three types of GWs signals (BH-BH in red, BH-NS in blue, NS-NS in yellow).}
	\label{fig:Cl_g0_YESTOMO}
\end{figure}
\begin{figure}[H]
    \centerline{			
		\subfloat{
			\includegraphics[width=1.0\linewidth]{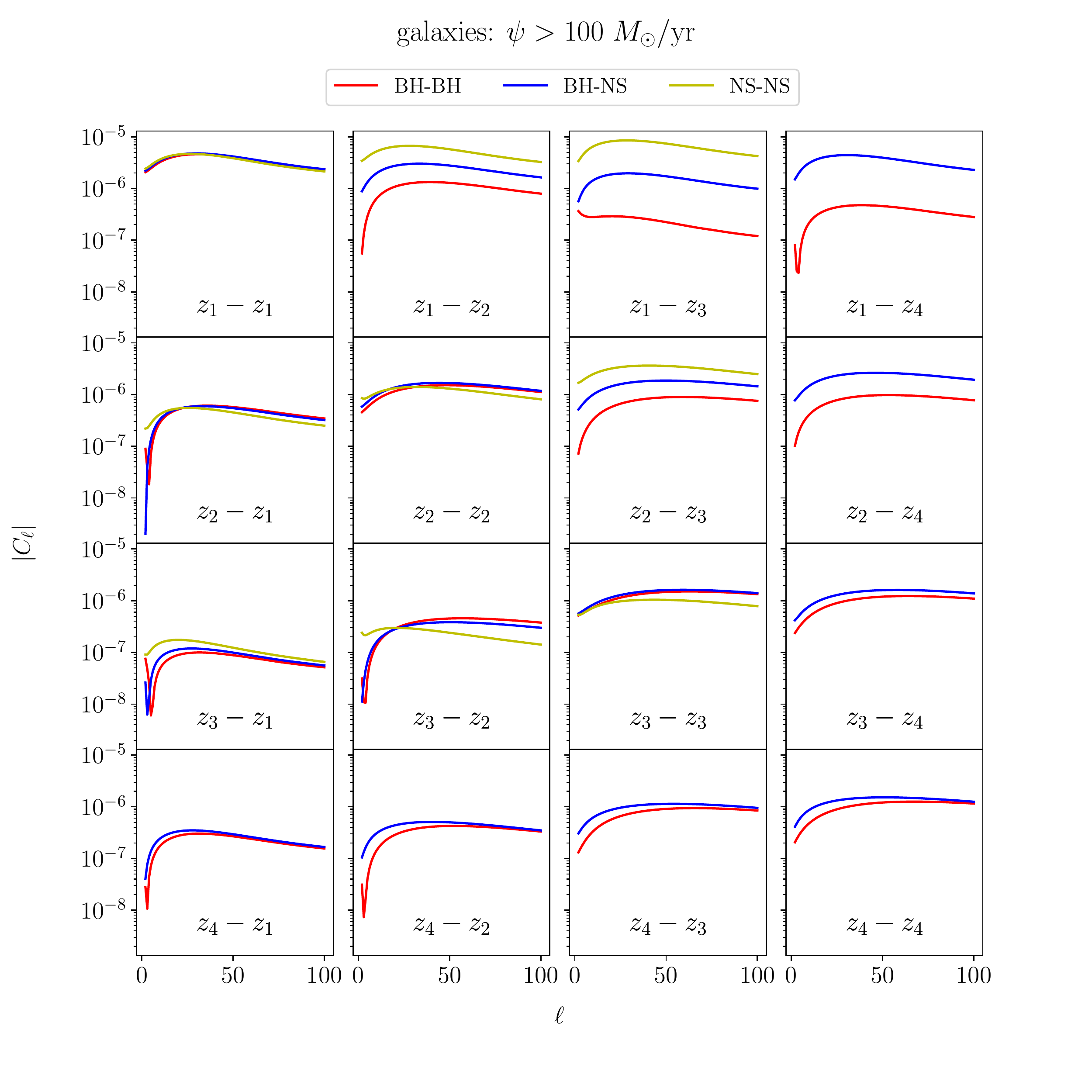}}}
	\caption{Angular power spectra $C_\ell$'s (absolute values) for the cross-correlations tomographic cases between galaxies with $\psi > 100 M_\odot/$yr with all three types of GWs signals (BH-BH in red, BH-NS in blue, NS-NS in yellow).}
	\label{fig:Cl_g1_YESTOMO}
\end{figure}
\begin{figure}[H]
    \centerline{
		\subfloat{
			\includegraphics[width=1.0\linewidth]{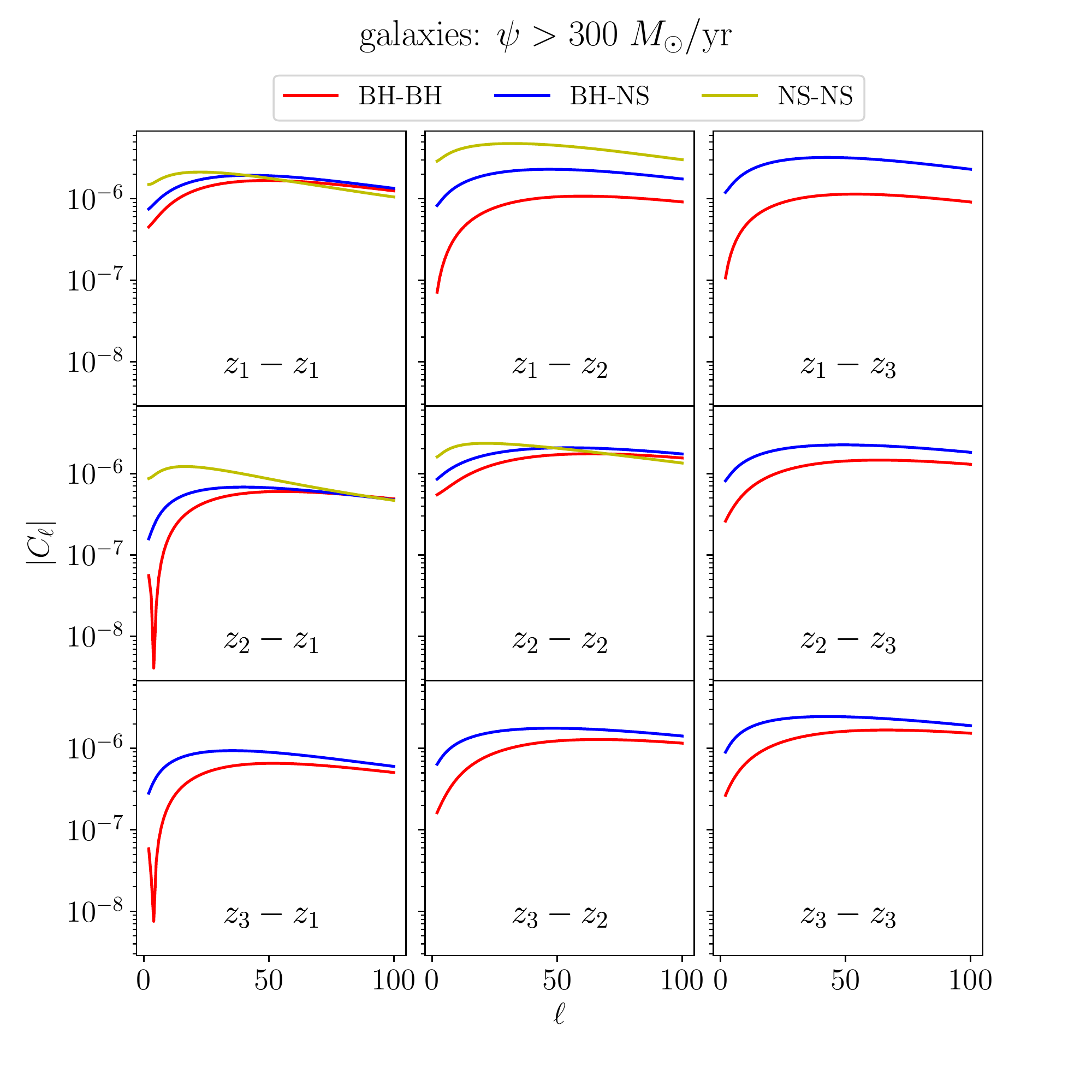}}}
	\caption{Angular power spectra $C_\ell$'s (absolute values) for the cross-correlations tomographic cases between galaxies with $\psi > 300 M_\odot/$yr with all three types of GWs signals (BH-BH in red, BH-NS in blue, NS-NS in yellow). }
	\label{fig:Cl_g2_YESTOMO}
\end{figure}

\bibliography{biblio}
\bibliographystyle{utcaps}

\end{document}